\documentclass[sigconf]{acmart}

\usepackage[utf8]{inputenc}
\usepackage[T1]{fontenc}
\usepackage{microtype}
\usepackage{color}
\usepackage{graphicx}
\usepackage{balance}  

\usepackage{subfigure}
\usepackage{multirow}
\usepackage{algorithm}
\usepackage{algorithmic}
\usepackage{color}
\usepackage{listings}
\usepackage{amsmath}
\usepackage{mathtools}
\usepackage{float}

\usepackage{amsmath}
\usepackage{subscript}
\usepackage[skip=0pt]{caption}
\usepackage[labelfont=bf]{caption}
\usepackage{ifthen}

\usepackage{graphicx} 
\newcommand{\eat}[1]{}

\newcommand{\showComments}{yes}
\newcommand{\submit}{yes} 

\newcommand{\note}[2]{
  \ifthenelse{\equal{\submit}{yes}}{}{%
    \ifthenelse{\equal{\showComments}{yes}}{\textcolor{#1}{#2}}{}
  }
}

\AtBeginDocument{%
  \providecommand\BibTeX{{%
    \normalfont B\kern-0.5em{\scshape i\kern-0.25em b}\kern-0.8em\TeX}}}
    
\setcopyright{acmcopyright}
\copyrightyear{2018}
\acmYear{2018}
\acmDOI{10.1145/1122445.1122456}

\acmConference[Woodstock '18]{Woodstock '18: ACM Symposium on Neural
  Gaze Detection}{June 03--05, 2018}{Woodstock, NY}
\acmBooktitle{Woodstock '18: ACM Symposium on Neural Gaze Detection,
  June 03--05, 2018, Woodstock, NY}
\acmPrice{15.00}
\acmISBN{978-1-4503-XXXX-X/18/06}

\begin{document}


\title{ A Comparison of Decision Forest Inference Platforms from A Database Perspective}



\newcommand{\tsc}[1]{\textsuperscript{#1}} 

\author{Hong Guan\tsc{1}, Mahidhar Reddy Dwarampudi*\tsc{1}, Venkatesh Gunda*\tsc{1}, Hong Min\tsc{2}, Lei Yu\tsc{3}, Jia Zou\tsc{1}}
\affiliation{
  \institution{\tsc{1} Arizona State University, \tsc{2} IBM T. J. Watson Research Center, \tsc{3} Rensselaer Polytechnic Institute}
  \country{}
}

\vspace{20pt}

\begin{abstract}
Decision forest, including RandomForest, XGBoost, and LightGBM, is one of the most popular machine learning techniques used in many industrial scenarios, such as credit card fraud detection, ranking, and business intelligence. 
Because the inference process is usually performance-critical, a number of frameworks were developed and dedicated for decision forest inference, such as ONNX, TreeLite from Amazon, TensorFlow Decision Forest from Google, HummingBird from Microsoft, Nvidia FIL, and lleaves. However, these frameworks are all decoupled with data management frameworks. \textit{It is unclear whether in-database inference will improve the overall performance.} In addition, these frameworks used different algorithms,  optimization techniques, and parallelism models. \textit{It is unclear how these implementations will affect the overall performance and how to make design decisions for an in-database inference framework.}
In this work, we investigated the above questions by comprehensively comparing the end-to-end performance of the aforementioned inference frameworks and netsDB, an in-database inference framework we implemented. Through this study, we identified that netsDB is best suited for handling small-scale models on large-scale datasets and all-scale models on small-scale datasets, for which it achieved up to hundreds of times of speedup. In addition, the relation-centric representation we proposed significantly improved netsDB's performance in handling large-scale models, while the model reuse optimization we proposed further improved netsDB's performance in handling small-scale datasets. 

\end{abstract}

\maketitle
\def\thefootnote{*}\footnotetext{These authors contributed equally to this work}\def\thefootnote{\arabic{footnote}}

\lstnewenvironment{DSL}
  {\lstset{
        aboveskip=5pt,
        belowskip=5pt,
        mathescape=true,
        basicstyle=\ttfamily\small,
        morekeywords={Set, Vector, Map, HashMap, bool, select,
  multiselect, aggregate, join, partition, member, method, construct, true, false, if, else, CREATE TABLE, PARTITION BY,
  self, literal, vector, return, for push_back, function, enum, sort, string, double},
        literate={~} {$\sim$}{1},
        showstringspaces=false}\vspace{0pt}%
   \noindent\minipage{0.47\textwidth}}
   {\endminipage\vspace{0pt}}

\lstnewenvironment{SQL}
  {\lstset{
        aboveskip=5pt,
        belowskip=5pt,
        escapechar=!,
        mathescape=true,
        basicstyle=\linespread{0.94}\ttfamily\small,
        morekeywords={JOIN, FROM, WHERE, SELECT},
        deletekeywords={VALUE, PRIOR},
        showstringspaces=false}
        \vspace{0pt}%
        \noindent\minipage{0.47\textwidth}}
  {\endminipage\vspace{0pt}}

\newcommand{\littlesection}[1]{\vspace{5pt}\noindent\textbf{#1}}
\section{Introduction}
\label{sec:intro}

Decision forest models, such as RandomForest~\cite{breiman2001random},   XGBoost~\cite{chen2016xgboost}, and LightGBM~\cite{ke2017lightgbm} are widely used machine learning algorithms for classification and regression tasks in production scenarios, such as search ranking in Microsoft Bing~\cite{ling2017model, zhu2017deep}, Alta Vista, Yahoo!~\cite{yin2016ranking}, and Yandex~\cite{styskin2011recency,lefortier2014online}, advertising in Facebook~\cite{he2014practical}, credit-card fraud prediction, healthcare, business intelligence. Compared to deep neural network models that are considered opaque, decision forest models are easier to identify and explain the significant variables in the data~\cite{arik2021tabnet}, which is important when applied to business decision processes that have compliance and audit requirements. They are popular for their robustness, scalability, and their abilities to handle a large number of features, handle missing data, and work well with both linear and non-linear relationships~\cite{kaggle-trend, dong2016characterizing, johnson2013learning, chapelle2011yahoo, qin2021neural}.  


Given the importance of decision forest models, there are many systems engineered recently to support and optimize the inference process of such models. Several highly cited examples of open-source systems include Scikit-Learn~\cite{pedregosa2011scikit}, ONNX~\cite{bai2019onnx}, TensorFlow Decision Forest (TFDF)~\cite{guillame2022yggdrasil}, TreeLite~\cite{cho2018treelite}, HummingBird~\cite{nakandala2020tensor}, lleaves~\cite{lleaves}, and Nvidia FIL~\cite{nvidia-fil, nvidia-fil-github}. 
However, we observed two significant gaps in these systems and related performance studies:

\noindent
\textbf{The data management gap.} All of these systems decouple the inference computation and data management and thus introduce additional latency for loading data from/to external data stores. 
Existing databases, such as IBM Infosphere Data Warehouse, ATLAS~\cite{wang2003atlas}, Oracle Data Mining, GLADE~\cite{Cheng2012GLADEBD}, MauveDB~\cite{mauvedb}, have support for “traditional ML” workloads based on Predictive Model Markup Language (PMML), stored procedures, user-defined functions (UDFs) and user-defined aggregates (UDAs), and views. However, there is little documentation discussing how these systems implement and optimize for decision forest workloads. \textit{It is unclear whether in-database inference will achieve lower end-to-end latency than the standalone inference platforms.} 

\begin{figure*}[]
\centering
\hspace{-10pt}
\subfigure[Naive Tree Traversal]{
\label{fig:naive-tree-traversal}
\includegraphics[width=0.33\textwidth]{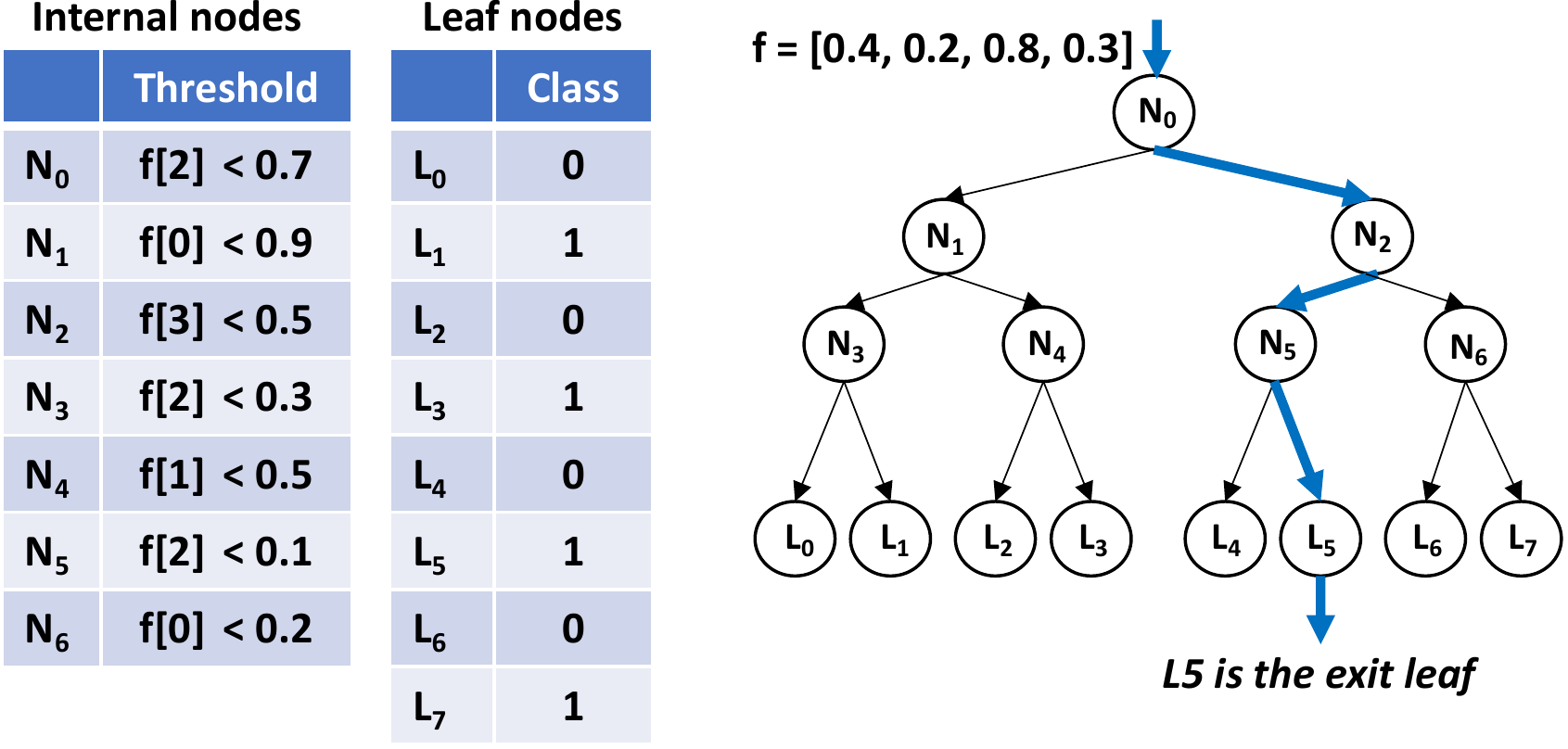}}
\subfigure[HummingBird]{
\label{fig:hummingbird}
\includegraphics[width=0.33\textwidth]{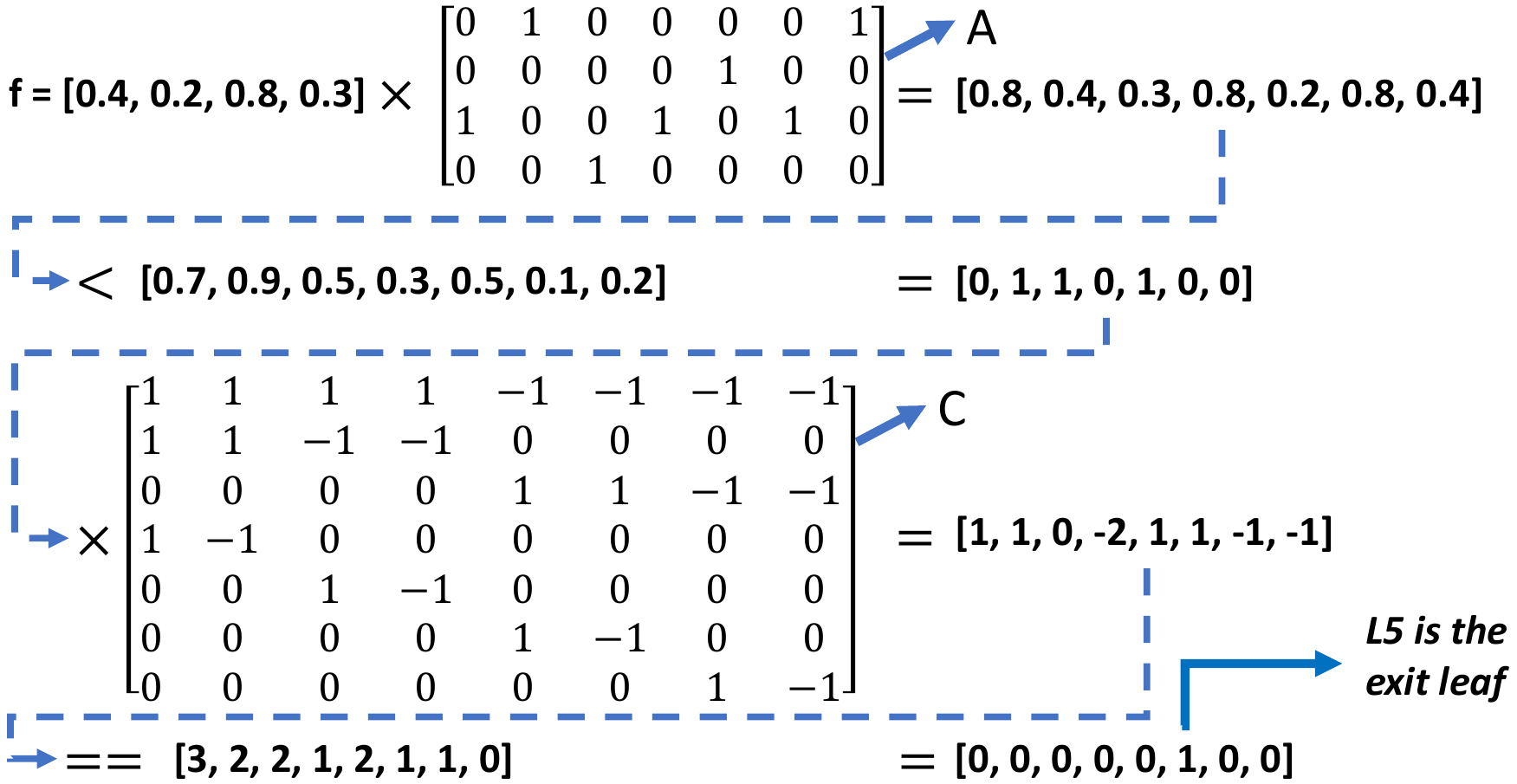}}
\subfigure[QuickScorer]{
\label{fig:quickscorer}
\includegraphics[width=0.33\textwidth]{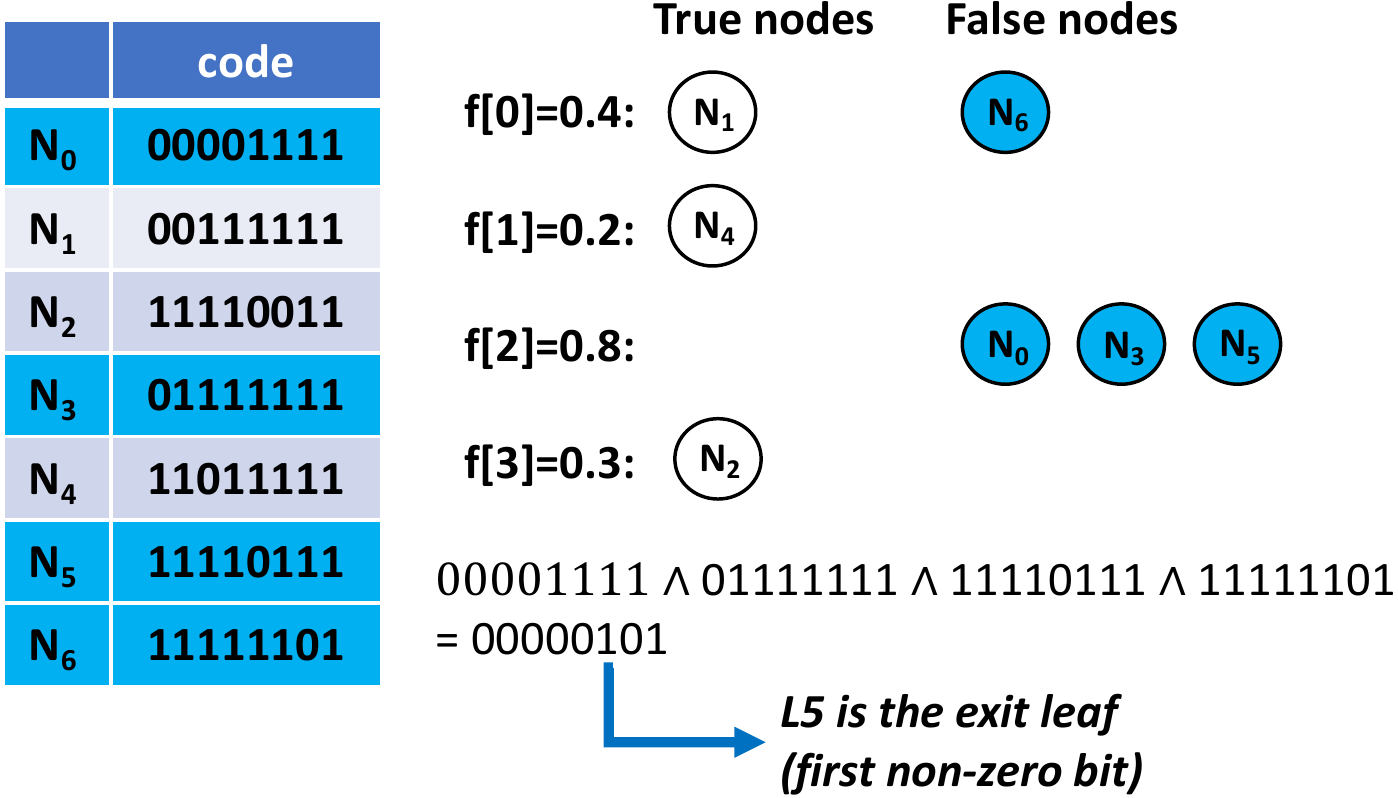}}
\caption{\label{fig:algorithms} \small
{Illustration of Decision Forest Algorithms.}
}
\end{figure*}

\begin{figure}[]
\centering
\hspace{-10pt}
\subfigure[Naive Prediction]{
\label{fig:naive-impl}
\includegraphics[width=0.26\textwidth]{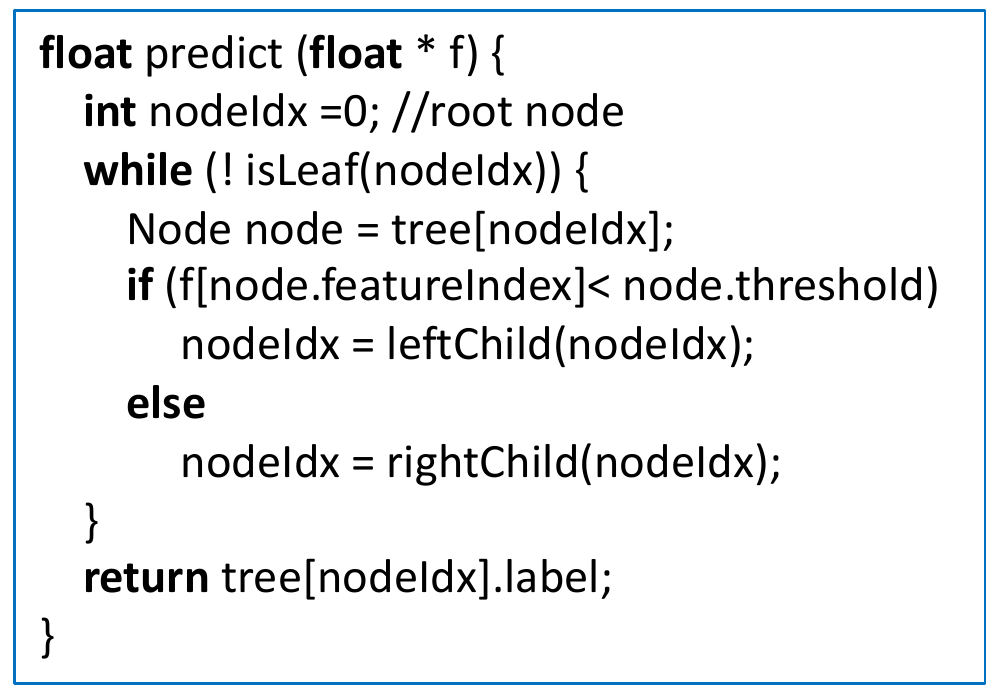}}
\subfigure[Compiled Prediction]{
\label{fig:compiled-impl}
\includegraphics[width=0.19\textwidth]{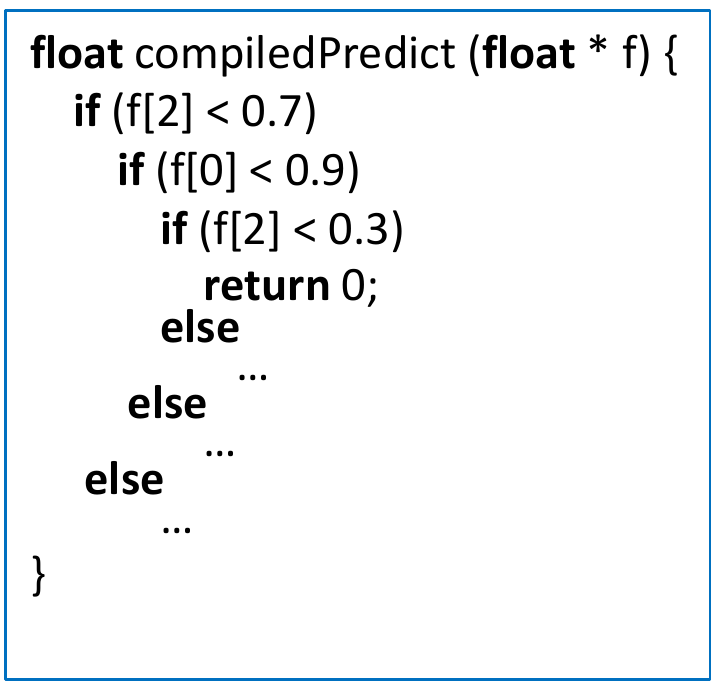}}
\subfigure[Predicated Prediction]{
\label{fig:predicated-impl}
\includegraphics[width=0.22\textwidth]{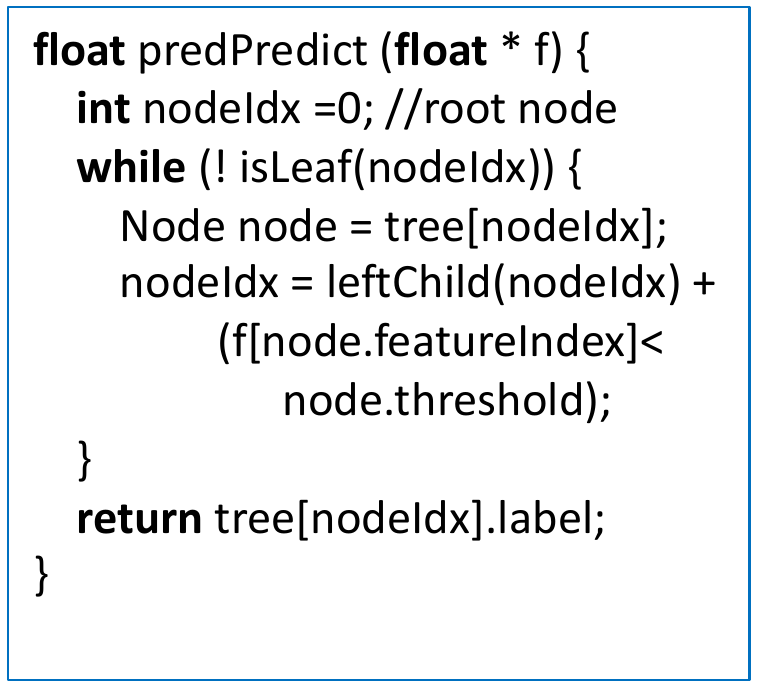}}
\subfigure[Vectorized Prediction]{
\label{fig:vectorized-impl}
\includegraphics[width=0.23\textwidth]{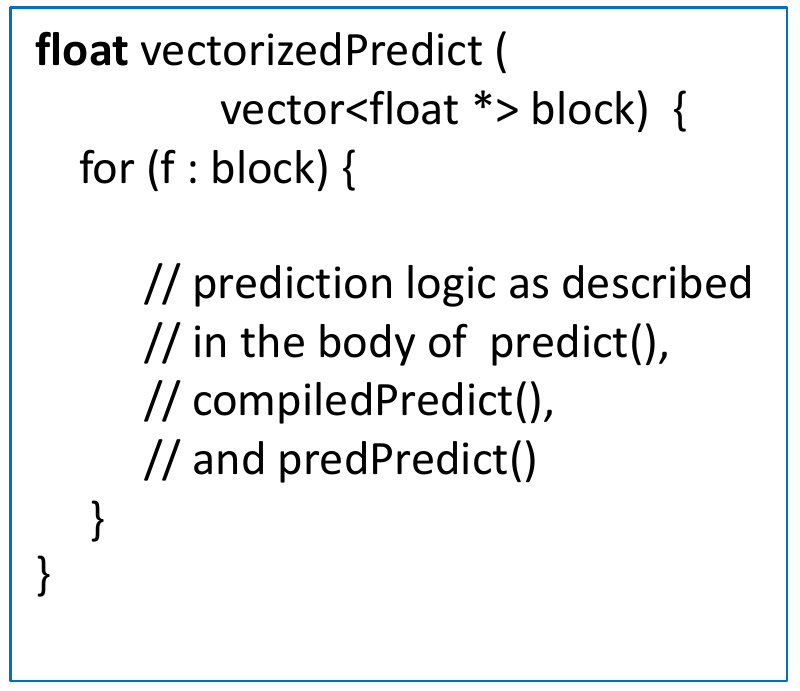}}
\caption{\label{fig:traversal-impl} \small
{Illustration of Different Tree-Traversal Implementations.}
}
\end{figure}

\vspace{6pt}
\noindent
\textbf{The performance understanding gap.} We identified key performance factors that cover the main architectural differences among the aforementioned decision forest inference frameworks. \textit{It is unclear how these design decisions will affect the end-to-end performance and how to make such design decisions for an in-database inference framework.} These key performance factors include:

\vspace{3pt}
\noindent
\textbf{F1. Algorithm.} There exist multiple algorithms for serving decision forest models, including: (1) \textbf{Naive tree traversal}, which traverses each decision tree from the root to the leaf, as illustrated in Fig.~\ref{fig:naive-tree-traversal} and Fig.~\ref{fig:naive-impl}. (2) \textbf{Compiled tree traveral}, which unrolls the naive tree traversal process into nested \texttt{if-else} blocks, as illustrated in Fig.~\ref{fig:compiled-impl}, to generate code that has conditional branches optimized~\cite{prasad2022treebeard, cho2018treelite, lleaves}. (3) \textbf{Predicated tree traveral}~\cite{asadi2013runtime}, which replaces conditional branches by predicates to avoid branch mis-predictions, as illustrated in Fig.~\ref{fig:predicated-impl}.
(4) \textbf{Hummingbird}~\cite{nakandala2020tensor}, which converts the decision tree inference into matrix computations, as illustrated in Fig.~\ref{fig:hummingbird} to leverage tensor computing libraries such as TVM~\cite{chen2018tvm}, TorchScript~\cite{Paszke_PyTorch_An_Imperative_2019}, and PyTorch~\cite{Paszke_PyTorch_An_Imperative_2019}. (5) \textbf{QuickScorer}~\cite{lucchese2017quickscorer, lucchese2015quickscorer}, which encodes each tree node into a bit vector and converts tree traversal into bit-wise AND operations of the bit vectors of all FALSE nodes (i.e., tree nodes that are evaluated as False for the current testing sample), as illustrated in Fig.~\ref{fig:quickscorer}. 

\vspace{3pt}
\noindent
\textbf{F2. Parallelism.} If using \textbf{data parallelism}, each thread runs the entire decision forest inference computation on a different data partition. When using \textbf{model parallelism}, each thread is responsible for running the inference of a partition of trees over the input data. Then the partial prediction results from all threads need to be aggregated to generate the final prediction.

\vspace{3pt}
\noindent
\textbf{F3. Batching.} Batching of testing samples for inferences is critical in balancing resource utilization and the overall latency. Different platforms and workloads desire different batch sizes to fully utilize the system resources. 

\vspace{3pt}
\noindent
\textbf{F4. Vectorization.}  Platforms such as lleaves, netsDB, Scikit-learn (only the RandomForest and XGBoost), and TFDF, applied vectorization, as illustrated in Fig.~\ref{fig:vectorized-impl}, to their inference functions, so that 
SIMD instructions can be leveraged to accelerate the inference. 

\subsection{Our Contributions} 

\noindent
\textbf{1. An in-database decision forest inference framework.}
To address the data management gap, we implemented a prototype of in-database decision forest inference in netsDB\footnote{https://github.com/asu-cactus/netsdb}, which is an object-oriented relational database management system (RDBMS) based on the PlinyCompute object model and query compilation framework implemented in C/C++ ~\cite{zou2018plinycompute, zou2019pangea, zou2020watson, zou2020architecture, DBLP:journals/pvldb/ZouDBIYJJ21, DBLP:journals/pvldb/ZhouCDMYZZ22, yuan2020tensor, jankov12declarative, shi2014mrtuner, jankov2020declarative}. 

We explored different representations of the decision forest inference in RDBMS. The \textbf{UDF-centric} representation encapsulates the entire inference computation into a single UDF. We found that when the size of the model 
increases, the cache locality worsens quickly using this approach. The \textbf{relation-centric} representation breaks down the inference computation into a flow of relational operators, including a \texttt{cross-product} operator that pairs up each decision tree and a block of testing samples. Then, for each pair, the prediction function from the decision tree is invoked to predict for each sample in the block. This operator is followed by an \texttt{aggregate} operator that aggregates all partial prediction results. 

Both representations support batching and vectorization. In addition, the UDF-centric representation adopts data parallelism so that each thread invokes the UDF on a data partition. The relation-centric representation adopts model parallelism, e.g., the cross-product operator is launched in multiple threads, with each thread responsible for a model partition (i.e., a subset of decision trees).
The \textbf{relation-centric} representation requires a model partitioning job stage. 
Once this job stage gets executed, its output can be reused for inferring different datasets. We thus proposed the model-reuse technique so that this job stage's output is materialized to accelerate executing different inference queries. 
%

\vspace{6pt}
\noindent
\textbf{2. A comprehensive performance comparison.} 
%
%
%
%
To address the performance understanding gap, we compared the dedicated decision forest inference frameworks and our proposed in-database inference framework on a broad class of workloads at all scales using RandomForest, XGBoost, and LightGBM. The types of datasets range from sparse to dense and narrow to wide. We compared the end-to-end latency of the decision forest inference process, including data loading, inference, and result writing.
The benchmarking framework is fully automated and open-sourced\footnote{https://github.com/asu-cactus/DFInferBench}.


\vspace{5pt}
\noindent
 \textbf{3. A series of interesting observations that may benefit future database and AI/ML systems design}, such as:

\noindent
$\bullet$ Data loading is a major performance bottleneck, particularly for (1) Workloads that infer large-scale datasets using small to medium forest models that have tens to hundreds of trees; and (2) Workloads that infer small-scale or wide-and-short datasets using all-scale models. For these workloads,  our in-database inference framework achieved up to hundreds of times speedup.

\eat{
\noindent
$\bullet$ For the two types of workloads mentioned above, CPU platforms usually achieved lower costs than GPU platforms. 
GPU platforms achieved significantly lower costs than CPU platforms when using large-scale forest models and large batches of data where inference rather than data loading becomes the major bottleneck.}

\noindent
$\bullet$ Cache locality is crucial to the performance of large-scale forest models. Therefore, model parallelism, which reduces the memory footprint of each thread/partition, outperformed data parallelism, where each thread needs to access the whole model in memory. 

\noindent
$\bullet$ The relation-centric representation significantly improved netsDB's performance in handling large-scale models, while the UDF-centric representation achieved the best performance for small-scale models. In addition, the model reuse optimization significantly improved the performance of the relation-centric representation for handling small-scale datasets.

\noindent
$\bullet$ GPU platforms achieved lower costs than CPU platforms for medium and large-scale forest models and large batches of data. But CPU platforms outperformed GPU platforms in other cases where data loading becomes the performance bottleneck.

\noindent
$\bullet$ If only inference computation time is considered, the QuickScorer algorithm achieved the best single-thread latency. The compilation-based approach combined with other optimizations, as implemented in lleaves, achieved the best inference latency among CPU platforms for large-scale (LightGBM) models. Nevertheless, such an approach required tremendous compilation latency, as detailed in Sec.~\ref{sec:model-conversion}.

\section{Survey of Existing Platforms}

\noindent
\textbf{A brief introduction about decision forest.}
RandomForest~\cite{breiman2001random}, XGBoost~\cite{chen2016xgboost}, and LightGBM~\cite{ke2017lightgbm} are three different decision forest training algorithms. Even training on the same dataset, with the same number of trees and the same maximum depth of each tree, the trained models could have different shapes using these training algorithms
~\cite{nakandala2020tensor}. 
The inference processes of the three decision forest models share the same first phase, which is to obtain the exit leaf for each tree in the forest. The second phase is slightly different. Taking Scikit-learn implementations of binary classification as an example, in the second phase, the RandomForest algorithm averages all trees' exit labels and then applies a sigmoid function to convert the averaged value to a probability score.  For XGBoost and LightGBM, the second phase sums up the weights of all exit labels and then obtains the final prediction using a sigmoid function. 

\vspace{6pt}
\noindent
\textbf{Platforms.} Due to the popularity and importance of the decision forest inference workloads, a number of platforms have been developed and dedicated for decision forest inferences. Next, we provide a survey about the most popular platforms:


\noindent
\textbf{Scikit-learn~\cite{pedregosa2011scikit}} It implements its own RandomForest algorithm but invokes the XGBoost~\cite{chen2016xgboost} and LightGBM~\cite{ke2017lightgbm} libraries to train and make inferences using XGBoost and LightGBM models. They all implement the naive tree traversal algorithm for inference, as illustrated in Fig.~\ref{fig:naive-tree-traversal}. 
Scikit-learn used model parallelism for random forest prediction, each thread runs the inferences of a partition of trees over the input data, and the results will be used to update a shared result vector protected by a lock. The predict function is vectorized by taking a batch of samples as input.
Both XGBoost and LightGBM libraries adopt data parallelism. The XGBoost library also uses vectorization, while the LightGBM does not.

\vspace{3pt}
\noindent
\textbf{ONNX~\cite{bai2019onnx} } It also uses the naive tree traversal algorithm. It chooses data parallelism or model parallelism based on the number of input samples and the number of trees in the forest. 
It does not exploit vectorization, and the underlying tree traversal function takes a single sample as input at a time.

\vspace{3pt}
\noindent
\textbf{HummingBird~\cite{nakandala2020tensor}} It transforms the tree traversal process into tensor computations, as illustrated in Fig.~\ref{fig:hummingbird}. It first converts the decision tree structure to two main tensors: (1) A tensor \texttt{A} that represents the relationships between each internal node and each feature; (2) A tensor \texttt{C} that captures the parent-child relationships among each pair of internal nodes. Then, the tensor of input samples is multiplied with tensor \texttt{A} to obtain the input path tensor. After that, the input path tensor is multiplied with tensor \texttt{C} to obtain the output path tensor. This way, existing tensor libraries on the CPU and GPU can be leveraged to accelerate the prediction process.

\vspace{3pt}
\noindent
\textbf{TensorFlow Decision Forest (TFDF)~\cite{tf-df}} It wraps a C++-based Yggdrasil library~\cite{guillame2022yggdrasil}, which implements the QuickScorer algorithm~\cite{lucchese2017quickscorer, lucchese2015quickscorer, ye2018rapidscorer} as well as the naive tree traversal algorithm. It will benchmark and select the best algorithm at the model compilation stage. The QuickScorer algorithm is illustrated in Fig.~\ref{fig:quickscorer}. It first encodes every tree node into a bit vector, with each bit corresponding to a leaf node. If the bit is set to 0, it means the leaf node is impossible to be the exit leaf if the current node is a FALSE node (i.e., evaluated to False). For a given input sample, the prediction of one decision tree can be obtained by applying the bit-wise AND operation to the bit vectors of all FALSE nodes. To identify FALSE nodes, the algorithm first groups the nodes from all trees in the model by features so that nodes regarding the same feature from different trees are stored together and sorted by their predicate thresholds. Given an input sample, it quickly determines all FALSE nodes using binary searches.

\vspace{3pt}
\noindent
\textbf{TreeLite~\cite{cho2018treelite}} It imports external models and partitions the trees into several compilation units. These compilation units will be transformed to C source functions in parallel. Each C source function corresponds to a compilation unit. As illustrated in Fig.~\ref{fig:compiled-impl}, it takes a single sample as input, runs a series of nested \texttt{if-else} blocks, and outputs the final predictions for trees belonging to the compilation unit. Then, these C source code functions will be further compiled into a shared library (i.e., a .so file)
. At prediction time, the shared library will be loaded to perform the prediction.

\vspace{3pt}
\noindent
\textbf{Nvidia FIL~\cite{nvidia-fil, nvidia-fil-github}} FIL implements a GPU version of the predicated tree traversal algorithm. Each GPU thread is responsible for inferring a batch of samples on one tree. To optimize the GPU cache locality, it exploits a reorganized dense tree representation in GPU memory, where the nodes at the same level but from different trees will be stored together to improve the cache locality. It also implements a sparse tree storage format, where the nodes from all trees are stored in one flat array. While nodes from one tree are stored together, sibling nodes that share the same parent node are always stored adjacently. 
For both cases, because the left child and right child are stored adjacently, it can use a predicate to replace the conditional branch. As illustrated in Fig.~\ref{fig:predicated-impl}, the conditional branch: \texttt{\textbf{if} (cond) \textbf{return}left(node\_idx) \textbf{else} \textbf{return} right(node\_idx)}, is replaced by \texttt{\textbf{return} left(node\_idx) + cond}, which avoided branches in GPU computation.

\vspace{3pt}
\noindent
\textbf{lleaves~\cite{lleaves}} It also compiles trees to nested \texttt{if-else} blocks. However, instead of translating the model into C source codes for compilation, lleaves designs an intermediate representation to describe the models and leverages the LLVM framework for code generation. Notably, lleaves is more optimized than TreeLite. For example, the functions generated by lleaves support vectorization. Lleaves currently only supports the LightGBM model on CPU.

\section{Our In-Database Inference Design} 
\label{sec:netsdb}

\begin{figure}[]
\centering
\hspace{-10pt}
\subfigure[UDF-Centric]{
\label{fig:udf-centric}
\includegraphics[width=0.09\textwidth]{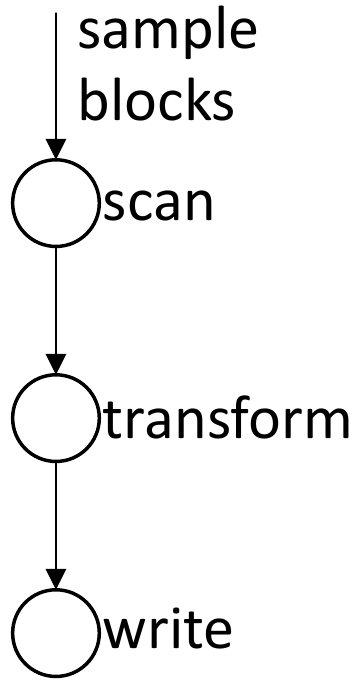}}
\subfigure[Relation-Centric]{
\label{fig:rel-centric}
\includegraphics[width=0.18\textwidth]{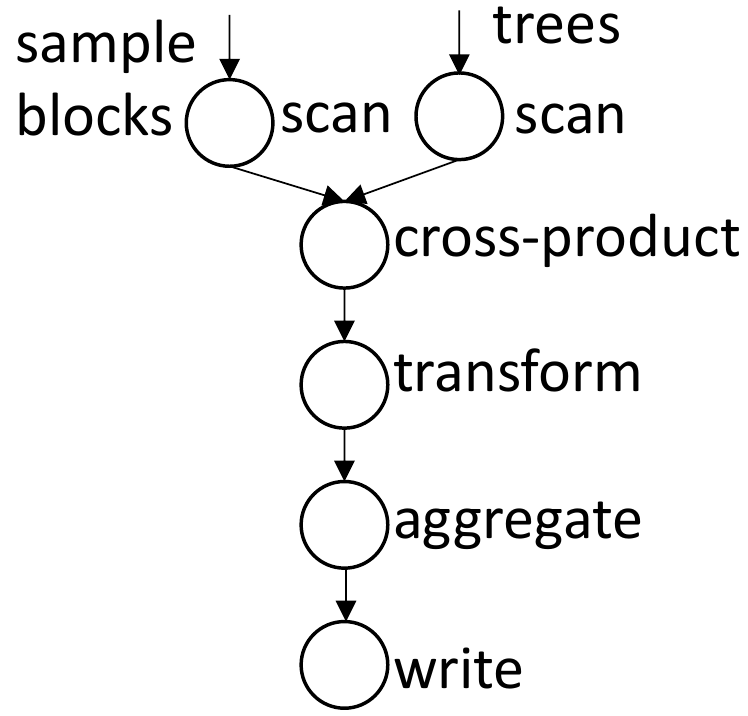}}
\subfigure[Model Reuse]{
\label{fig:model-reuse}
\includegraphics[width=0.18\textwidth]{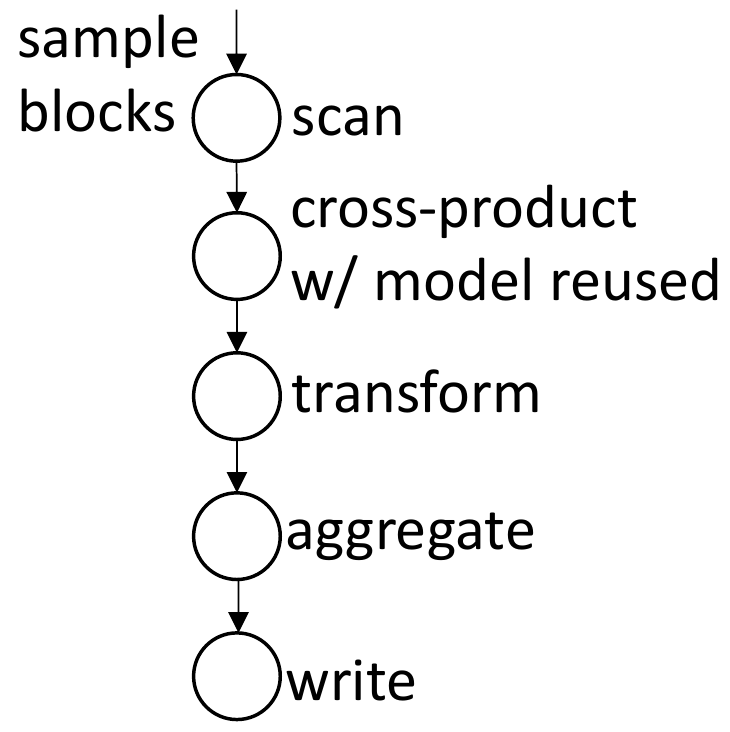}}
\caption{\label{fig:database-representations} \small
{Computation graph for various database representations.}
}
\end{figure}

\subsection{Key Design Decisions}
\label{sec:db-decisions}
\noindent
\textbf{Algorithm.} We found that the naive tree traversal algorithm and the predicated tree traversal algorithm are best suited for RDBMS. The compiled tree traversal algorithm requires code generation for UDFs such as the predict function, which is not supported in many RDBMS. The HummingBird algorithm relies on optimization for expensive matrix computations, and it is not suitable for most RDBMS designed for CPU platforms. The QuickScorer algorithm is hard to be represented in a relation-centric style. One reason is that the QuickScorer model groups all nodes by features; thus, the model is essentially a collection of feature groups. Because the sizes of feature groups are not well-balanced, it is challenging to partition the model evenly. In addition, C++ lacks a good library for large bit vectors that have more than $64$ bits, which limits the number of leaf nodes to $64$ in a decision tree. For example, the TFDF's implementation of QuickScorer is single-threaded and only supports a maximum tree depth of $6$ (i.e., $log_2^{64}$). 
We thus implemented both naive tree traversal and predicated tree traversal algorithms in netsDB, but we did not observe any significant difference in the inference time for most of the workloads.

 \vspace{3pt}
\noindent
\textbf{Storage of Input Samples.} Physically, the input samples are stored as a collection of tensor blocks, called sample blocks. Each block is a 2D tensor that represents a vector of feature vectors.

\vspace{3pt}
\noindent
\textbf{Batching.} The application can specify a batch size to control the number of sample blocks in one batch. Then, the application can iteratively issue an inference query to process one batch until all batches are processed. 

\vspace{3pt}
\noindent
\textbf{Scheduling and Parallelism.} In netsDB, similar to other dataflow frameworks such as Apache Spark~\cite{zaharia2010spark, armbrust2015spark}, the users develop their applications as dataflow graphs, where every node represents a relational operator (which should be customized using UDFs) and every edge represents a dataset. At runtime, a dataflow graph is split into multiple pipeline (job) stages. A pipeline stage is a series of atomic operators (e.g., \texttt{scan}, \texttt{transform}, \texttt{hash}, \texttt{partition}, \texttt{probe}, \texttt{join}, \texttt{cross-product}, \texttt{aggregate}, \texttt{write}) with the last operator being a pipeline breaker that materializes the output data (e.g., \texttt{hash}, \texttt{partition}, \texttt{aggregate}, \texttt{write}). For example, a \texttt{cross-product} relational operator will be split into multiple atomic operators belonging to different pipeline stages as described in Sec.~\ref{sec:rel-centric}. Each pipeline stage is executed with multiple threads. These threads run the same logic but with different input data. Usually, we configure the number of threads for each pipeline stage to be the number of CPU cores. As we will detail in Sec.~\ref{sec:udf-centric} and Sec.~\ref{sec:rel-centric}, the UDF-centric representation of decision forest inference implements data parallelism, and the relation-centric representation implements model-parallel inferences.

\vspace{3pt}
\noindent
\textbf{Vectorization.} When executing a pipeline stage, each thread iteratively fetches a vector of sample blocks and runs a series of atomic operators to process the vector. Each atomic operator in the pipeline stage is vectorized. It takes in a vector of sample blocks and outputs a vector of result blocks to the next atomic operator. 

\subsection{The UDF-Centric Representation} 
\label{sec:udf-centric}
With the UDF-centric representation, the decision forest inference logic is encapsulated in a single UDF that customizes a \texttt{transform} operator (i.e., like a \texttt{map} function). The UDF contains a forest object, which is a vector of trees, and each tree is stored as a vector of tree nodes. The UDF has a prediction function that takes a sample block as input and outputs a block of predictions. The prediction function iteratively processes each sample in the block by traversing each tree and aggregating the prediction of all trees. 

As illustrated in Fig.~\ref{fig:udf-centric}, the corresponding dataflow graph for a simple example of UDF-centric inference consists of a \texttt{transform}  operator that is customized by a UDF, which takes a sample block as input and outputs a result block.  After compilation, the dataflow graph is scheduled as one pipeline stage. Each thread iteratively fetches a vector of sample blocks, runs the prediction UDF over it, and writes the final predictions to an output dataset. The UDF is initialized by parsing a pre-trained scikit-learn model. 

The benefit of this approach is in its simplicity, e.g., the inference process is compiled to a single pipeline stage. The encapsulation also facilitates extending the UDF to invoke functions from popular libraries, including GPU libraries. 
The shortcoming is that each CPU core needs to access the whole forest model, which leads to significant cache misses for large-scale models. 

\subsection{The Relation-Centric Representation}
\label{sec:rel-centric}

To reduce the cache misses for inferences with large-scale models, we propose a relation-centric representation that facilitates model parallelism.
As illustrated in Fig.~\ref{fig:rel-centric}, we represent the decision forest inference process as a dataflow graph that uses two key operators:
(1) A \texttt{cross-product} operator, which performs a Cartesian product between the collection of trees and the collection of input sample blocks, and enumerates all possible pairs of tree and sample block.
Each pair is further converted into a block of prediction results (e.g., the return class or weight associated with the exit leaf)  using a \texttt{transform} UDF that performs tree traversal over each sample in the block iteratively. 
(2) An \texttt{aggregate} operator is used to aggregate all prediction results for the same sample using different aggregation logic for different types of models.

\noindent
To reduce the cache misses, we implement the \texttt{cross-product} operator in a model-parallel fashion. It partitions the model into many subsets of decision trees and allocates each partition to a thread. Then, a pointer to each page from the inference dataset is sent to all threads for generating the cross-product and partial prediction results. Using this approach, each thread only needs to access a partition of trees. It significantly reduces the cache misses and shortens the inference latency for large-scale models, as observed in our experiments.  

\noindent
\textbf{Model Reuse.} While the relation-centric representation works well for inferring large datasets, it brings significant overheads for processing small datasets. That is because compared to the UDF-centric representation that only requires one pipeline stage, the relation-centric representation is compiled to multiple pipeline stages because there exist multiple pipeline breakers in the dataflow graph. For example, one stage partitions the model, one stage runs the \texttt{cross-product} and groups partial prediction results by tree IDs, one stage aggregates the prediction results from all trees, and one stage post-processes the aggregated result (e.g., applying the sigmoid function) and writes the final output to a dataset. To reduce the overheads, we identified that the model-partitioning stage can be reused by different inference applications as long as they use the same model. Therefore, we can materialize the results of this pipeline stage and directly reuse the materialized results to simplify the dataflow graph, as illustrated in Fig.~\ref{fig:model-reuse}.

\section{Benchmark Workload Description}

We evaluated the performance of various decision forest models at all scales on well-known classification datasets, as shown in Tab.~\ref{tab:datasets}.

\begin{table}[h]
\centering
\scriptsize
\caption{\label{tab:datasets} {\small Statistics of the Datasets}}
\begin{tabular}{|l|r|r|l|r|} \hline
Dataset&NumRows&NumFeatures&Prediction Type&Testing Ratio\\\hline \hline
Epsilon&$100$K&$2000$&Classification&$20\%$\\ \hline
Fraud&$285$K &$28$&Classification&$20\%$\\ \hline
Year&$515$K&$90$&Regression&$20\%$ \\ \hline
Bosch&$1.184$M&968&Classification&$20\%$\\ \hline
Higgs&$11$M&$28$&Classification&$20\%$ \\ \hline
Criteo&51M&1M&Classification&$11\%$\\ \hline
Airline&$115$M&$13$&Classification/Regression&$20\%$\\ \hline
TPCx-AI (SF=30)&131M&7&Classification&$100\%$\\ \hline
\end{tabular}
\end{table}

Most of the platforms investigated in this work, such as ONNX, HummingBird, TreeLite, lleaves, and netsDB, can not directly train a model. Therefore, we use Scikit-learn to train models using RandomForest, XGBoost, and LightGBM algorithms on each evaluation dataset in Tab.~\ref{tab:datasets}. For each type of model, we used a different number of trees, ranging from $10$ to $1,600$, with each tree having a maximum depth of $8$. These are all widely used hyper-parameters~\cite{nakandala2020tensor}. We then convert these models to be loaded to each platform for inferences. The performance of the model conversion and loading process is discussed in Sec.~\ref{sec:model-conversion}.

For most datasets except TPCx-AI~\cite{tpcx-ai} and Criteo~\cite{criteo}, we used $80\%$ of samples to train models. Then we convert the trained model to run inferences against the $20\%$ remaining samples on each target platform. For TPCx-AI, we used the fraud detection scenario~\cite{tpcx-ai}. We trained the model on a smaller dataset with scale factor (SF) 1, then tested the model on the dataset with SF 30, which is described in Tab.~\ref{tab:datasets}. For Criteo, the training and testing data was pre-split by the data provider~\cite{libsvmdata,criteo}, as illustrated in Tab.~\ref{tab:datasets}.

\noindent
\textbf{Target Scenarios.} In this work, we focused on the end-to-end latency of inference pipelines that ran all types of models to make inferences for datasets that are managed by native or external data stores. In practice, features are extracted from databases in many industrial scenarios, such as fraud detection and recommendation based on transaction records~\cite{lam2021automated}.
Therefore, for netsDB, the testing datasets were stored natively. For other platforms, the testing datasets, except for Epsilon and Criteo, were stored in tabular format in a PostgreSQL database installed on the same machine, with the database connection accelerated using the state-of-art ConnectorX library~\cite{wang2022connectorx}. Besides, Epsilon has $2000$ features, and Criteo has $1$ million features, but  PostgreSQL only supports up to $1600$ columns~\cite{postgresql-limits}. Therefore, for Epsilon, we stored each tuple in Array type in PostgreSQL, which turned out to be slower than the tabular format in data loading  as detailed in Sec.~\ref{sec:epsilon}. For Criteo, we simply loaded it from a LIBSVM file in sparse storage format~\cite{CC01a}.

We measured data loading time, inferences time, and data writing time in an end-to-end inference process. We did not consider the model conversion and model loading time as part of the end-to-end latency because these times can be amortized to multiple inference queries. We will discuss such one-time costs in Sec.~\ref{sec:model-conversion}.
\eat{
}

\eat{
\vspace{5pt}
\noindent
\textbf{Model Conversion.} Target systems, such as ONNX, HummingBird, TreeLite, lleaves, and Nvidia FIL, all support importation from Sciki-learn models. TFDF only supports the conversion for RandomForest models. With the help of Google engineers, we developed our own XGBoost converter for TFDF.
}

\vspace{-10pt}
\section{Experimental Configuration}
\label{sec:config}
\noindent
\textbf{Software Configuration.} We used version Scikit-learn v1.1.2, ONNX v1.12.0, Hummingbird v0.4.5, Nvidia Rapids-FIL v22.08, TreeLite v2.3.0, TFDF v0.2.7, PostgreSQL v14. For lleaves, netsDB, and ConnectorX, we used the code download from their Github master repositories. For all platforms, we carefully tune the number of threads to fully use the computational resources.

We ran the HummingBird models in several different backends, including Pytorch v1.13.1,  TorchScript (within Pytorch v1.13.1), and TVM v0.10.0,  with and without GPU acceleration. 

Nvidia FIL provides multiple options for inferences: (1) Auto, which automatically estimates and selects the best-performing strategy; (2) Batch Tree Reorganization for the dense forest storage format; (3) Using the sparse forest storage format. We used the Auto option by default. 

The TFDF platform only supported the loading from Scikit-learn RandomForest models~\cite{tf-df}. With the help of Google TFDF engineers, we developed a converter to import XGBoost models to TFDF models. To this point, there is no API available for parallelizing the TFDF inference process, and it is slower than other platforms in most cases, and we omit the TFDF results in the overall evaluation (Sec.~\ref{sec:overall-evaluation}) and only discuss them in the detailed analysis (Sec.~\ref{sec:analysis}).

\noindent
\textbf{Hardware Configuration.}
For CPU experiments, we used the AWS EC2 r4.2xlarge instance with $8$ CPU cores and $62$ gigabytes of memory. All instances are installed with Linux Ubuntu $20$ and $200$ gigabytes SSD storage. The cost of the instance is \$$0.532$ per hour. 
For the GPU experiments, we used an AWS g4dn.2xlarge instance, which has an NVIDIA T4 Tensor Core GPU with $16$ gigabytes memory and an eight-core CPU with $32$ gigabytes host memory. Its cost is \$$0.752$ per hour.

To resolve the performance variations in cloud environments, we repeated each experiment multiple times. In total, we used more than $\textit{\textbf{10,000}}$ hours of the AWS EC2 platform for the evaluation.

\noindent
\textbf{Profiling}
We monitored the various system resource utilization for CPU, memory, and disk I/O. In addition, we used Linux Perf to profile the elapsed time as well as cache misses. We used gpustat, a wrapper built on nvidia-smi, for profiling the GPU performance and memory utilization.
\section{Overall Evaluation Results}
\label{sec:overall-evaluation}
We first describe the overall evaluation results in three categories: small-scale dense datasets, medium to large-scale dense datasets, and sparse and/or wide datasets. For each case/measurement, the batch size is carefully tuned for each platform to achieve the lowest latency. We will discuss how batch size affects the overall performance in the detailed analysis (Sec.~\ref{sec:analysis}). Finally, we will discuss the model conversion and loading overheads which are not counted in the overall evaluation results.

\subsection{Small-Scale Dense Datasets}
In this section, we will summarize the results on two relatively smaller datasets: Fraud and Year, as described in Tab.~\ref{tab:datasets}. 

The overall benchmark results for Fraud and Year are illustrated in Tab.~\ref{tab:fraud} and Tab.~\ref{tab:year}. 
It showed that among all CPU/GPU platforms, netsDB with the UDF-centric representation (netsDB-UDF) achieved the lowest latency for small models that have $10$ trees. NetsDB with the model reuse optimization (netsDB-OPT) significantly reduced the latency for the relation-centric representation (netsDB-Rel) and achieved the lowest latency for medium to large models with $500$ and $1600$ trees across all platforms. That is because inference on such small datasets is significantly faster than data transfer, and data transfer thus becomes the major bottleneck. This bottleneck gets avoided using in-database inference.

\begin{table*}[]
\begin{center}
\small
\resizebox{\textwidth}{!}{
\begin{tabular}{|c||c|c|c|c|c|c|c|c|c|c||c|c|c|c|}
\hline
 &\multicolumn{10}{|c||}{CPU} & \multicolumn{4}{c|}{GPU}\\
&Sklearn&ONNX&HB-Pytorch&HB-TS&HB-TVM&TreeLite&lleaves&netsDB-UDF&netsDB-Rel&netsDB-OPT&HB-Pytorch&HB-TS&HB-TVM&FIL\\ 
\hline
\multicolumn{15}{|c|}{\texttt{RandomForest}} \\ 
\hline
10 Trees  &1.0&1.0&1.0&1.3&1.0&1.0&-&\textbf{0.5}&2.1&0.6&\textbf{0.9}&1.4&\textbf{0.9}&1.0\\
500 Trees  &1.3&1.1&4.4&4.0&1.5&1.2&-&1.3&2.3&\textbf{0.7}&1.0&1.4&\textbf{0.9}&1.0\\
1600 Trees  &2.2&1.4&13.0&11.0&2.8&1.8&-&3.5&2.6&\textbf{1.0}&1.2&1.6&\textbf{1.0}&\textbf{1.0}\\
\hline
\multicolumn{15}{|c|}{\texttt{XGBoost}} \\ 
\hline
10 Trees  &1.0&1.0&1.0&1.3&1.0&1.0&-&\textbf{0.5}&2.1&0.7&\textbf{0.9}&1.7&\textbf{0.9}&1.0\\
500 Trees  &1.1&1.1&4.5&4.1&1.7&1.2&-&1.0&2.3&\textbf{0.8}&1.0&1.7&\textbf{0.9}&1.0\\
1600 Trees  &1.3&1.4&13.0&11.0&2.5&1.4&-&2.1&2.6&\textbf{1.0}&1.2&1.7&\textbf{1.0}&\textbf{1.0}\\
\hline
\multicolumn{15}{|c|}{\texttt{LightGBM}} \\ 
\hline
10 Trees  &1.0&1.0&1.0&1.4&1.0&1.0&1.0&\textbf{0.5}&2.1&0.6&\textbf{0.9}&1.7&\textbf{0.9}&1.0\\
500 Trees  &1.5&1.2&4.4&4.0&1.7&1.5&1.0&1.1&2.3&\textbf{0.8}&1.0&1.7&\textbf{0.9}&1.0\\
1600 Trees  &2.5&1.6&12.8&10.7&2.5&2.6&1.2&1.7&2.8&\textbf{1.1}&1.2&1.7&\textbf{0.9}&1.0\\
\hline
\end{tabular}
}
\caption{End-to-End Latency Comparison for Fraud. (Unit: seconds)}
\label{tab:fraud}
\end{center}
\end{table*}

\begin{table*}[]
\begin{center}
\small
\resizebox{\textwidth}{!}{
\begin{tabular}{|c||c|c|c|c|c|c|c|c|c|c||c|c|c|c|}
\hline
 &\multicolumn{10}{|c||}{CPU} & \multicolumn{4}{c|}{GPU}\\
&Sklearn&ONNX&HB-Pytorch&HB-TS&HB-TVM&TreeLite&lleaves&netsDB-UDF&netsDB-Rel&netsDB-OPT&HB-Pytorch&HB-TS&HB-TVM&FIL\\ 
\hline
\multicolumn{15}{|c|}{\texttt{RandomForest}} \\ 
\hline
10 Trees  &4.7&4.7&4.8&5.1&4.7&4.7&-&\textbf{0.5}&2.3&0.8&\textbf{3.8}&4.3&\textbf{3.8}&3.9\\
500 Trees  &5.5&5.2&11.5&10.4&5.9&6.4&-&1.4&2.9&\textbf{1.3}&\textbf{3.9}&4.4&\textbf{3.9}&4.0\\
1600 Trees  &7.3&6.5&27.4&23.1&7.7&10.8&-&4.8&4.2&\textbf{2.6}&4.3&4.6&\textbf{4.0}&\textbf{4.0}\\
\hline
\multicolumn{15}{|c|}{\texttt{XGBoost}} \\ 
\hline
10 Trees  &4.6&4.7&4.7&5.2&4.7&4.7&-&\textbf{0.5}&2.3&0.8&\textbf{3.8}&4.6&\textbf{3.8}&3.9\\
500 Trees  &5.1&5.1&11.2&10.3&6.2&6.1&-&1.6&2.8&\textbf{1.2}&4.0&5.0&\textbf{3.8}&3.9\\
1600 Trees  &6.1&5.9&26.7&23.2&7.1&9.7&-&4.8&3.8&\textbf{2.3}&4.3&5.3&\textbf{4.0}&\textbf{4.0}\\
\hline
\multicolumn{15}{|c|}{\texttt{LightGBM}} \\ 
\hline
10 Trees  &4.7&4.7&4.8&5.1&4.7&4.7&4.7&\textbf{0.5}&2.3&0.8&\textbf{3.8}&4.3&\textbf{3.8}&3.9\\
500 Trees  &6.3&5.2&11.3&10.3&6.1&6.1&5.0&1.5&2.7&\textbf{1.2}&4.0&4.4&\textbf{3.8}&4.0\\
1600 Trees  &10.2&6.1&26.5&22.7&7.4&9.9&5.9&5.0&3.7&\textbf{2.2}&4.3&4.6&\textbf{4.0}&4.1\\
\hline
\end{tabular}
}
\caption{End-to-End Latency Comparison for Year. (Unit: seconds)}
\label{tab:year}
\end{center}
\end{table*}

\eat{
\begin{table}[]
\begin{center}
\small
\begin{tabular}{|c||c|c||c|c|}\hline
&\multicolumn{2}{|c||}{Fraud} & \multicolumn{2}{c|}{Year}\\
&CPU&GPU&CPU&GPU\\ 
\hline
\multicolumn{5}{|c|}{\texttt{RandomForest}} \\ 
\hline
10 Trees  &\textbf{0.01} &0.02 &\textbf{0.01}&0.08 \\
500 Trees  &\textbf{0.01} &0.02 &\textbf{0.02}&0.08 \\
1600 Trees  &\textbf{0.01}&0.02&\textbf{0.04}&0.08\\
\hline
\multicolumn{5}{|c|}{\texttt{XGBoost}} \\ 
\hline
10 Trees  &\textbf{0.01} &0.02 &\textbf{0.01}&0.08 \\
500 Trees  &\textbf{0.01} &0.02 &\textbf{0.02}&0.08 \\
1600 Trees  &\textbf{0.01}&0.02&\textbf{0.03}&0.08\\
\hline
\multicolumn{5}{|c|}{\texttt{LightGBM}} \\ 
\hline
10 Trees  &\textbf{0.01} &0.02 &\textbf{0.01}&0.08 \\
500 Trees  &\textbf{0.01} &0.02 &\textbf{0.02}&0.08 \\
1600 Trees  &\textbf{0.02} &\textbf{0.02} &\textbf{0.03}&0.08\\
\hline
\end{tabular}
\caption{Comparison of CPU cost for the best-performed CPU platform on r4.2xlarge (\$$0.532$ per hour), and GPU cost for the best-performed GPU platform on g4dn.2xlarge (GPU, \$$0.752$ per hour) for inferring on small datasets (Unit:cents)}
\label{tab:small-cpu-gpu}
\end{center}
\end{table}
}

\begin{figure}[h]
\centering
\vspace{-10pt}
\subfigure[10 trees]{
\label{fig:10trees}
\includegraphics[width=0.48\textwidth]{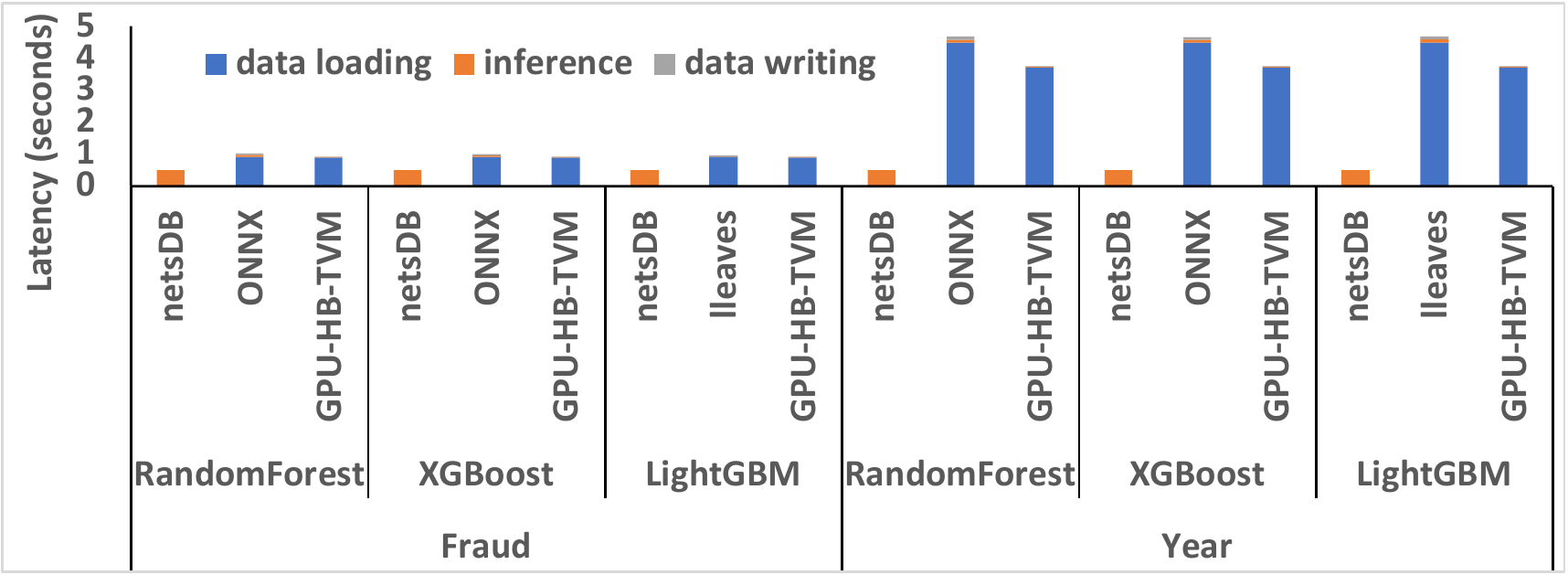}}
\subfigure[500 trees]{
\label{fig:500trees}
\includegraphics[width=0.48\textwidth]{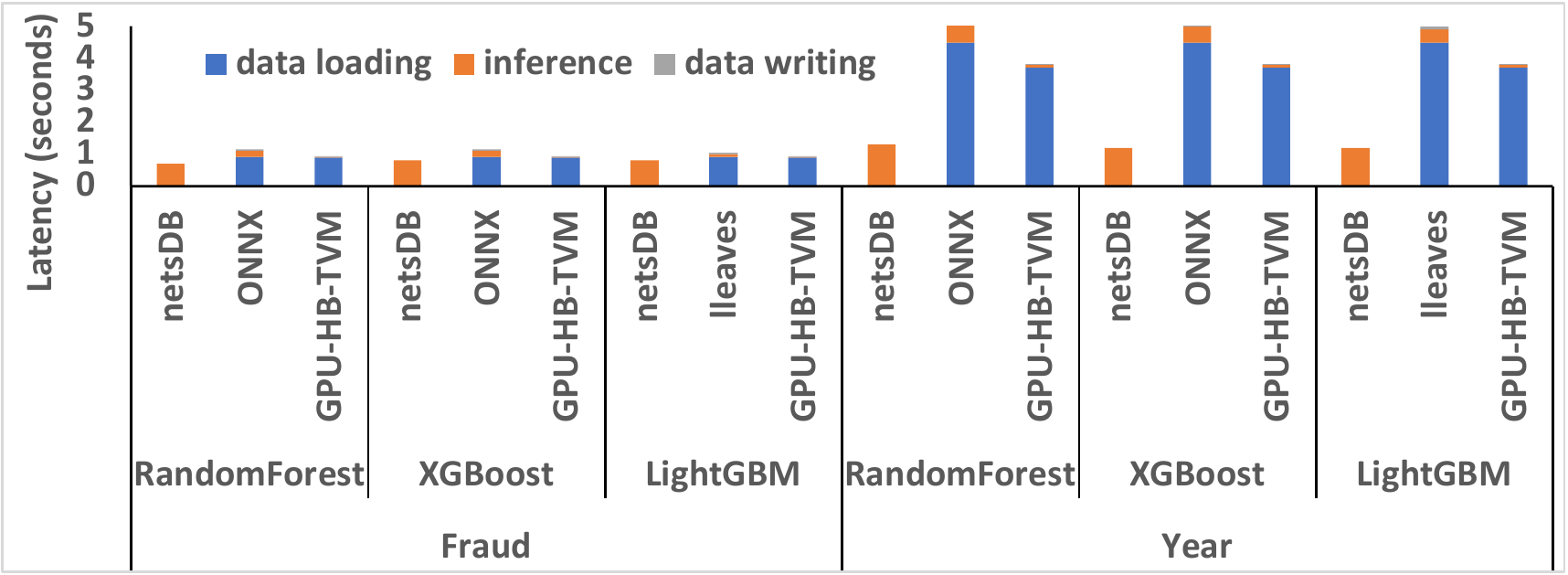}}
\subfigure[1600 trees]{
\label{fig:1600trees}
\includegraphics[width=0.48\textwidth]{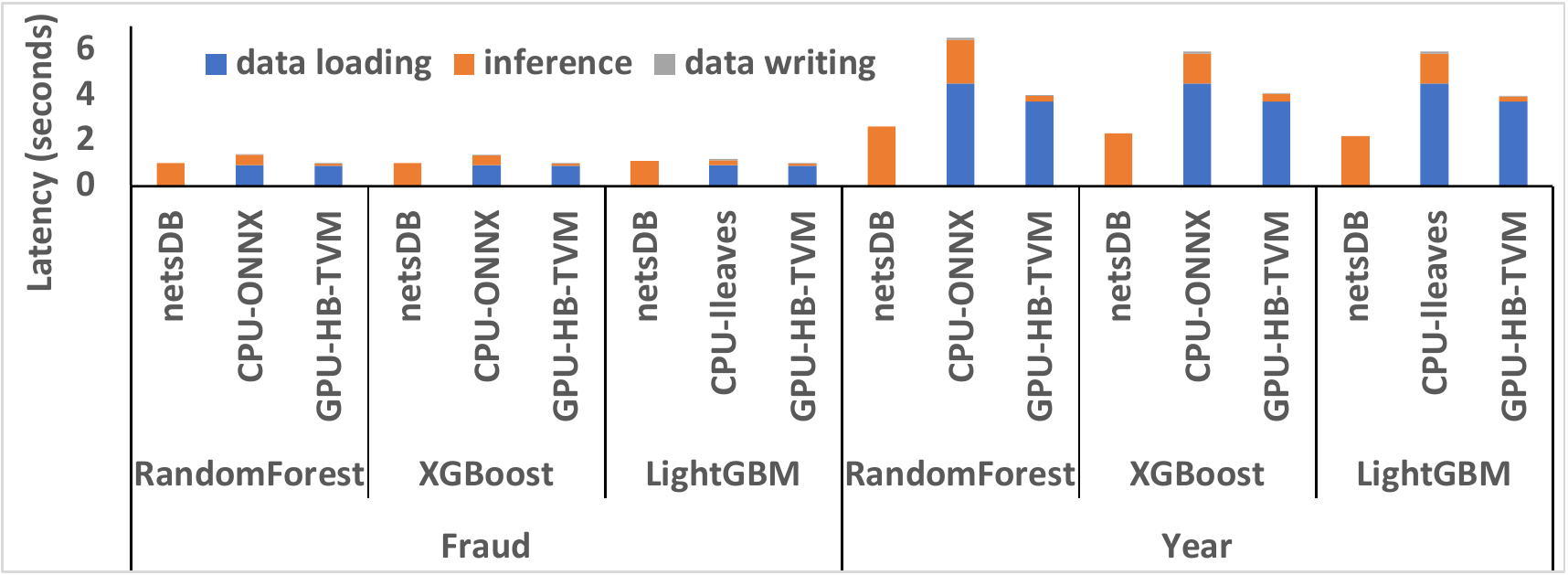}}
\caption{\label{fig:small-breakdown} \small
{\color{black}Latency breakdown for netsDB, the fastest of the rest CPU platforms, and the fastest GPU platform on small datasets.}
}
\end{figure}

\subsubsection{Fraud} 

We have the following key observations on this workload.
First, netsDB outperformed other CPU/GPU platforms. That is because data transfer is the major bottleneck of the Fraud workload, as illustrated in Fig.~\ref{fig:small-breakdown}. When applying $10$-tree models to the Fraud dataset, data transfer accounts for $90\%$ of overall latency on ONNXCPU, $95\%$ on lleaves, and $97\%$ on HB-TVM. However, the ratio of data transfer latency to the overall latency decreases with the increase in model sizes. For example, for $1600$-tree models, the aforementioned ratios dropped to $65\%$, $76\%$, and $88\%$ respectively.
Second, netsDB-Rel performed significantly worse than other CPU/GPU platforms for models with $500$ and $1,600$ trees. That is mainly because it runs multiple pipeline stages, including partitioning the model to prepare for the \texttt{cross-product} operation. The scheduling and materialization overheads are significant compared to the inference latency.  The model reuse technique, as in netsDB-OPT, resolved the issue and improved the performance.
Third, netsDB, ONNX, and lleaves achieved better monetary costs than GPU platforms (see Sec.~\ref{sec:config} for AWS cost information).
It indicates that GPU may not be very helpful for the inferences on small datasets, which is ubiquitous in real-time applications where testing samples are batched in small-size buffers to guarantee low buffering latency. 
%

%


\subsubsection{Year} 
The observations for the Year workload are similar to Fraud, except that netsDB achieved significantly higher speedups on the Year dataset. 
For example, as illustrated in Tab.~\ref{tab:year}, for small models with $10$ trees, netsDB-UDF achieved $7.6\times$ speedup compared to the best GPU platform, and $9.4\times$ speedup compared to the best of the rest CPU platform. But the corresponding speedups achieved on the Fraud dataset were merely $1.8\times$ and $2\times$, respectively. 
Then, for $500$-tree models, netsDB-OPT achieved $3\times$ speedup compared to the best GPU platform and $4\times$ speedup compared to the second-best CPU platforms, while the corresponding speedups achieved on the Fraud dataset was merely $1.1$ to $1.3\times$ for CPU and $1.1$ to $1.5\times$ for GPU. 
In addition, for the case of $1,600$-tree models, for the Year dataset, netsDB-OPT achieved $1.7$ to $2.5\times$ speedup compared to the best GPU platform and $1.5$ to $2.7\times$ speedup compared to the second-best CPU platforms. However, for the same case, there was no significant speedup achieved on the Fraud dataset by netsDB-OPT compared to GPU platforms, and the speedup compared to the second-best CPU platforms was merely $1.1$ to $1.4\times$.

As illustrated in Fig.~\ref{fig:small-breakdown}, the data loading and writing times accounted for $96\%$, $96\%$, and $98\%$ on ONNXCPU, lleaves, and HB-TVM, respectively, which were even higher than the corresponding ratios for the Fraud workload. This explained the better speedup achieved on Year. Similar to Fraud, the ratio of data loading latency to the overall end-to-end latency decreased with the increase in model sizes. For $1600$-tree models, the aforementioned ratios dropped to $69\%$, $77\%$, and $94\%$ for the Year case.

\subsubsection{Summary of Findings}
The key findings are as follows:

\noindent
(1) The in-database inferences outperformed all CPU/GPU platforms for small datasets because the data loading process is always a major performance bottleneck.

\noindent
(2) NetsDB achieved lower costs than GPU platforms in all cases, and GPU is not very helpful for the inferences over small datasets, which are insufficient to fully utilize GPU resources. The inference computation time is insufficient to compensate for the overheads of transferring data between CPU and GPU.


\subsection{Medium to Large-Scale Dense Datasets}

This section mainly investigates the performance of three workloads,  Higgs, Airline, and TPCx-AI. The overall benchmark results for these datasets are illustrated in Tab.~\ref{table:higgs}, Tab.~\ref{table:airline}, and Tab.~\ref{table:tpcx-ai}. When processing those larger-scale datasets, the latency difference between netsDB-Rel and netsDB-OPT is insignificant compared to the overall latency, so we omitted netsDB-OPT in the results.
\begin{table*}[]
\begin{center}
\small
\resizebox{\textwidth}{!}{
\begin{tabular}{|c||c|c|c|c|c|c|c|c|c||c|c|c|c|}
\hline
 &\multicolumn{9}{|c||}{CPU} & \multicolumn{4}{c|}{GPU}\\
&Sklearn&ONNX&HB-Pytorch&HB-TS&HB-TVM&TreeLite&lleaves&netsDB-UDF&netsDB-Rel&HB-Pytorch&HB-TS&HB-TVM&FIL\\ 
\hline
\multicolumn{14}{|c|}{\texttt{RandomForest}} \\ 
\hline
10 Trees  &8.3&9.3&8.5&8.9&8.5&7.8&-&\textbf{0.9}&2.7&\textbf{7.7}&8.6&7.8&8.4\\
500 Trees  &28.2&20.3&60.6&48.4&29.3&42.4&-&24.4&\textbf{14.2}&11.7&10.6&9.4&\textbf{8.7}\\
1600 Trees  &70.8&45.2&205.5&167.1&80.8&130.4&-&98.6&\textbf{42.0}&20.6&16.1&\textbf{11.0}&11.1\\
\hline
\multicolumn{14}{|c|}{\texttt{XGBoost}} \\ 
\hline
10 Trees  &8.1&9.3&8.7&8.8&8.8&8.3&-&\textbf{0.6}&2.3&8.0&8.6&\textbf{7.8}&8.0\\
500 Trees  &19.1&20.8&57.5&47.5&33.0&35.3&-&25.9&\textbf{13.1}&11.6&11.0&8.4&\textbf{8.3}\\
1600 Trees  &35.9&37.6&Failed&Failed&61.5&99.1&-&98.51&\textbf{34.7}&20.3&16.1&10.1&\textbf{8.8}\\
\hline
\multicolumn{14}{|c|}{\texttt{LightGBM}} \\ 
\hline
10 Trees  &8.5&9.3&8.8&9.1&8.8&8.5&8.1&\textbf{0.9}&2.7&\textbf{7.8}&8.6&\textbf{7.8}&8.2\\
500 Trees  &39.6&18.9&56.8&47.6&32.8&34.6&14.5&26.4&\textbf{14.2}&11.6&11.1&\textbf{8.5}&\textbf{8.5}\\
1600 Trees  &113.2&39.2&202.8&183.3&61.1&102.8&\textbf{29.5}&119&38&20.3&16.1&9.1&\textbf{9.0}\\
\hline
\end{tabular}
}
\caption{End-to-End Latency Comparison for Higgs.  (Unit: seconds)}
\label{table:higgs}
\end{center}
\end{table*}
%

\begin{table*}[]
\begin{center}
\small
\resizebox{\textwidth}{!}{
\begin{tabular}{|c||c|c|c|c|c|c|c|c|c||c|c|c|c|}
\hline
 &\multicolumn{9}{|c||}{CPU} & \multicolumn{4}{c|}{GPU}\\
&Sklearn&ONNX&HB-Pytorch&HB-TS&HB-TVM&TreeLite&lleaves&netsDB-UDF&netsDB-Rel&HB-Pytorch&HB-TS&HB-TVM&FIL\\ 
\hline
\multicolumn{14}{|c|}{\texttt{RandomForest}} \\ 
\hline
10 Trees  &60.7&74.2&77.5&71.8&63.5&55.1&-&\textbf{3.3}&17.6&45.7&45.7&\textbf{45.3}&46.9\\
500 Trees  &210.6&146.2&1502.2&1229.5&273.1&336.3&-&302.3&\textbf{82.4}&86.0&70.8&52.0&\textbf{50.3}\\
1600 Trees  &543.5&305.5&5206.3&4144.7&760.8&1052.0&-&1120.6&\textbf{239.4}&177.7&126.4&70.6&\textbf{62.5}\\
\hline
\multicolumn{14}{|c|}{\texttt{XGBoost}} \\ 
\hline
10 Trees  &59.2&73.3&77.9&69.8&63.8&59.8&-&\textbf{2.9}&18.6&45.7&45.8&\textbf{45.3}&46.1\\
500 Trees  &143.4&137.8&1468.9&1212.8&305.2&264.7&-&272.3&\textbf{80.1}&85.3&70.7&51.3&\textbf{47.5}\\
1600 Trees  &340.2&287.1&5130.3&4232.3&613.1&862.0&-&1071.1&\textbf{220.0}&175.8&125.0&68.4&\textbf{51.9}\\
\hline
\multicolumn{14}{|c|}{\texttt{LightGBM}} \\ 
\hline
10 Trees  &60.5&74.7&77.5&71.8&64.3&59.4&58.8&\textbf{2.8}&18.6&45.7&45.8&\textbf{45.3}&46.7\\
500 Trees  &339.7&144.1&1421.2&1142.6&306.4&224.5&98.8&96.9&\textbf{76.5}&85.2&70.9&51.4&\textbf{49.8}\\
1600 Trees  &1039.2&296.5&4909.6&4198.1&614.0&654.8&\textbf{185.0}&914.8&220.7&176.1&125.1&68.5&\textbf{59.5}\\
\hline
\end{tabular}
}
\caption{End-to-End Latency Comparison for Airline. (Unit: seconds)}
\label{table:airline}
\end{center}
\end{table*}

\begin{table*}[]
\begin{center}
\small
\resizebox{\textwidth}{!}{
\begin{tabular}{|c||c|c|c|c|c|c|c|c|c||c|c|c|c|}
\hline
 &\multicolumn{9}{|c||}{CPU} & \multicolumn{4}{c|}{GPU}\\
&Sklearn&ONNX&HB-Pytorch&HB-TS&HB-TVM&TreeLite&lleaves&netsDB-UDF&netsDB-Rel&HB-Pytorch&HB-TS&HB-TVM&FIL\\ 
\hline
\multicolumn{14}{|c|}{\texttt{RandomForest}} \\ 
\hline
10 Trees  &449.4&513.3&462.6&469.9&448.3&450.1&-&\textbf{3.2}&56.2&444.5&442.6&\textbf{442.3}&448.2\\
500 Trees  &1233.9&865.1&4521.1&3087.0&1635.4&1061.0&-&1391.7&\textbf{363.7}&678.3&584.6&480.7&\textbf{465.6}\\
1600 Trees  &3386.0&1649.3&15302.0&10482.5&4311.0&3153.7&-&5000.7&\textbf{1551.4}&1201.4&906.5&605.7&\textbf{526.7}\\
\hline
\multicolumn{14}{|c|}{\texttt{XGBoost}} \\ 
\hline
10 Trees  &438.4&485.5&438.4&441.9&461.3&441.5&-&\textbf{2.3}&22.2&444.6&\textbf{443.1}&447.6&444.6\\
500 Trees  &925.7&883.4&3821.6&2967.8&1835.4&1560.1&-&1328.1&\textbf{369.9}&678.3&583.5&476.6&\textbf{451.9}\\
1600 Trees  &2014.7&1614.1&14093.3&11509.2&3590.2&4928.4&-&4735.7&\textbf{1612.9}&1193.0&895.1&578.6&\textbf{479.6}\\
\hline
\multicolumn{14}{|c|}{\texttt{LightGBM}} \\ 
\hline
10 Trees  &445.1&509.9&434.1&436.7&447.0&434.2&434.7&\textbf{2.8}&22.2&443.3&\textbf{442.8}&445.7&447.7\\
500 Trees  &2069.0&916.7&3711.0&3023.6&1876.2&1070.9&586.9&1122.0&\textbf{366.2}&672.9&580.5&478.4&\textbf{463.3}\\
1600 Trees  &6146.0&1739.3&12196.5&11143.7&3646.8&3388.8&\textbf{997.2}&4623.6&1599.5&1186.2&890.1&573.6&\textbf{529.5}\\
\hline
\end{tabular}
}
\caption{End-to-End Latency Comparison for TPCx-AI. (Unit: seconds).}
\label{table:tpcx-ai}
\end{center}
\end{table*}

\eat{
\begin{table}[]
\begin{center}
\small
\begin{tabular}{|c||c|c||c|c||c|c|}\hline
&\multicolumn{2}{|c||}{Higgs} & \multicolumn{2}{c||}{Airline}& \multicolumn{2}{c|}{TPCx-AI}\\
&CPU&GPU&CPU&GPU&CPU&GPU\\ 
\hline
\multicolumn{7}{|c|}{\texttt{RandomForest}} \\ 
\hline
10 Trees  &\textbf{0.01} &0.16 &\textbf{0.05}&0.94&\textbf{0.05} &9.24 \\
500 Trees  &0.21 &\textbf{0.18} &1.22&\textbf{1.05}&\textbf{5.38} &9.73 \\
1600 Trees  &0.62 &\textbf{0.23}&3.53&\textbf{1.31}&22.93 &\textbf{11.0}\\
\hline
\multicolumn{7}{|c|}{\texttt{XGBoost}} \\ 
\hline
10 Trees  &\textbf{0.01} &0.16 &\textbf{0.04}&0.95&\textbf{0.03}&9.26\\
500 Trees  &0.19 &\textbf{0.17} &1.18&\textbf{0.99}&\textbf{5.46}&9.44\\
1600 Trees  &0.51 &\textbf{0.18} &3.25&\textbf{1.08}&23.84&\textbf{10.02}\\
\hline
\multicolumn{7}{|c|}{\texttt{LightGBM}} \\ 
\hline
10 Trees  &\textbf{0.01} &0.16 &\textbf{0.04}&0.95&\textbf{0.04}&9.25\\
500 Trees  &0.21 &\textbf{0.18} &1.13&\textbf{1.04}&\textbf{5.41}&9.68\\
1600 Trees  &0.44 &\textbf{0.19} &2.73&\textbf{1.24}&14.73&\textbf{11.06}\\
\hline
\end{tabular}
\caption{Comparison of CPU cost of the best-performed CPU platform on r4.2xlarge (\$$0.532$ per hour), and GPU cost of the best-performed GPU platform on g4dn.2xlarge (\$$0.752$ per hour) for medium to large-scale datasets (Unit:cents)}
\label{table:large-cpu-gpu}
\end{center}
\end{table}
}

\begin{figure*}[]
\centering
\vspace{-10pt}
\subfigure[Higgs]{
\label{fig:higgs-breakdown}
\includegraphics[width=1\textwidth]{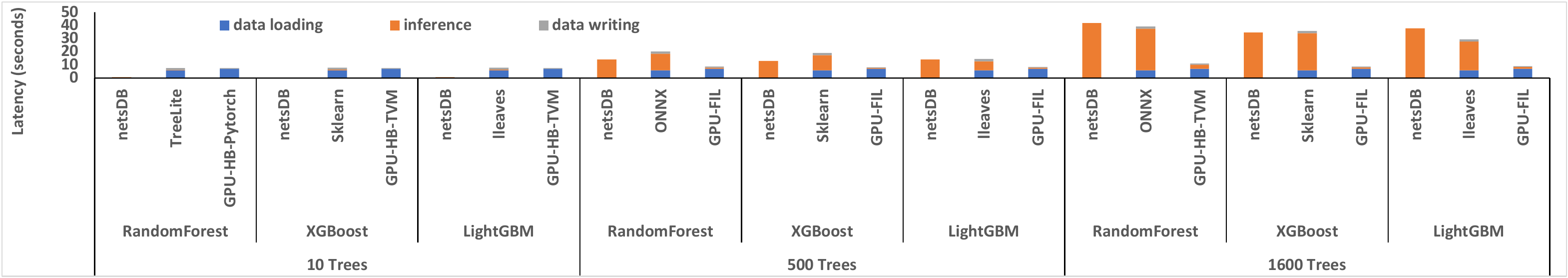}}
\subfigure[Airline]{
\label{fig:airline-breakdown}
\includegraphics[width=1\textwidth]{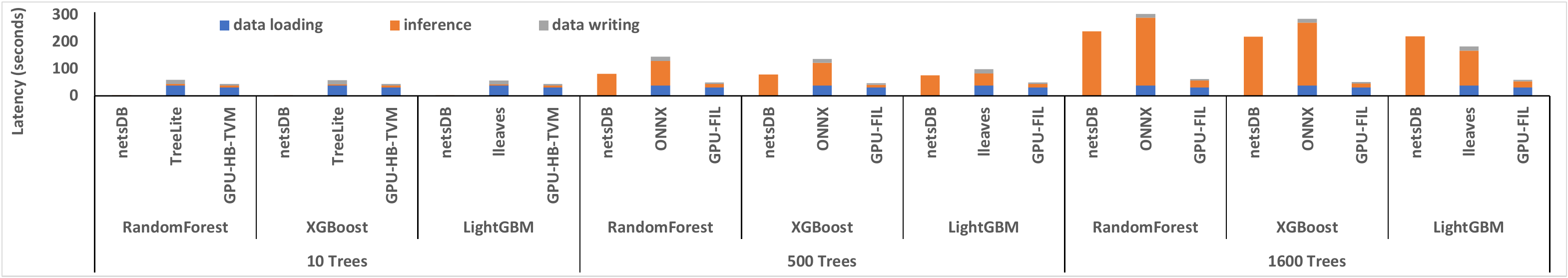}}
\subfigure[TPCx-AI]{
\label{fig:tpcx-ai-breakdown}
\includegraphics[width=1\textwidth]{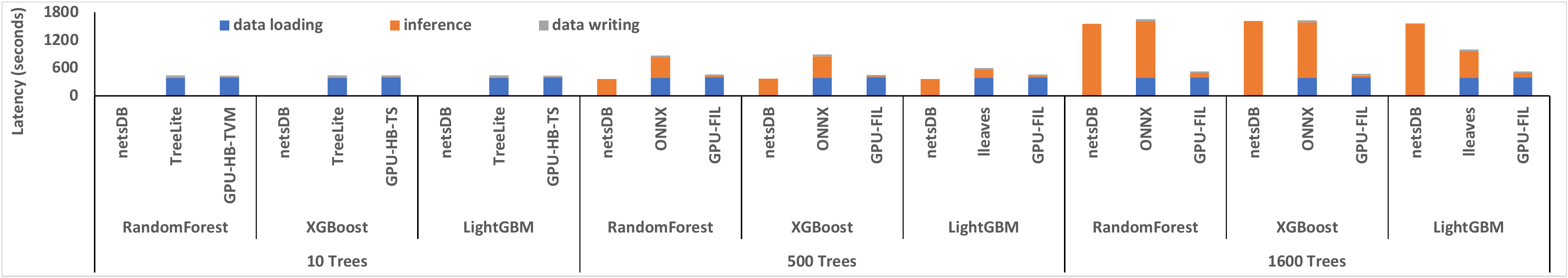}}
\caption{\label{fig:large-breakdown} \small
{\color{black}Latency breakdown for the netsDB, the fastest of the rest CPU platforms, and the fastest GPU platform for  medium to large datasets.}
}
\end{figure*}

\subsubsection{Higgs} 

As illustrated in Tab.~\ref{table:higgs}, when only using CPU processors, for a $10$-trees model, netsDB-UDF achieved $8$ to $10$ times speedup compared to the fastest of the rest CPU platforms. When using $500$ trees, netsDB-Rel achieved $1.3\times$ to $1.6\times$ speedup compared to the fastest of the rest CPU platforms. For large-scale decision forest models with $1,600$ trees, netsDB is slightly worse than the lleaves for LightGBM, but it is still slightly better than the Scikit-learn (XGBoost) and ONNX platforms and significantly better than all other CPU platforms.
When serving $500$ to $1,600$ trees, utilizing GPU significantly accelerated the decision forest prediction at even lower costs. Nvidia FIL and HummingBird with TVM as the backend achieved the best end-to-end performance on GPU.

As illustrated in Fig.~\ref{fig:higgs-breakdown}, most of the performance gain when using netsDB was contributed by the avoidance of the data transfer time, which is the major bottleneck for serving small-scale models. However, when the number of trees increases to $1600$ trees,   inference, instead of data transfer, becomes the primary bottleneck, as illustrated in Fig.~\ref{fig:higgs-breakdown}. This explained the drop in the performance gain brought by netsDB with the increase in model sizes.

\subsubsection{Airline} 
As illustrated in Tab.~\ref{table:airline}, the speedup achieved by netsDB on this workload is even higher than the Higgs workload. In terms of end-to-end time, for $10$-tree models, the netsDB-UDF achieved $18\times$-$21\times$ speedup compared to the fastest of the rest of the CPU platforms. For $500$-tree models, netsDB-Rel achieved $1.3\times$ to $1.9\times$ speedup compared to the fastest of the rest CPU platforms. Similar to Higgs, the performance gain shrinks with the increase in model size. For RandomForest and XGBoost $1,600$-tree models, netsDB-Rel achieved $1.3\times$ speedup. For LightGBM with $1,600$ trees, lleaves still performed slightly better than netsDB. In addition, similar to Higgs, the GPU platforms achieved significantly lower monetary costs than the CPU platforms in most cases. 

As illustrated in Tab.~\ref{tab:datasets}, the size of the Airline testing dataset is five times larger than Higgs. As a result, the time spent in data transfer accounts for a significantly higher proportion in the end-to-end latency than Higgs, as observed in Fig.~\ref{fig:airline-breakdown}. This explained the increased performance gain of netsDB for the Airline case.

\vspace{-5pt}
\subsubsection{TPCx-AI} This workload used all $131$ million of samples for inferences. 
With the increase in input dataset size,  the data transfer time accounts for an even higher proportion in the overall latency than Higgs and Airline, as illustrated in Fig.~\ref{fig:large-breakdown}. 
Correspondingly, the speedup achieved by netsDB also increased compared to Higgs and Airline, as illustrated in Tab.~\ref{table:tpcx-ai}.  

NetsDB-UDF achieved more than $100 \times$ speedup compared to all other CPU/GPU platforms for the $10$-tree models. Moreover, for models with $500$ trees, netsDB-Rel achieved more than $1.6\times$ to $2.4\times$ speedup compared to the fastest of other CPU platforms, and it achieved $1.2\times$ to $1.3\times$ speedup compared to the fastest of the GPU platforms. Even for $1,600$-tree models, netsDB-Rel outperformed all other CPU platforms for RandomForest and XGBoost, though it is slower than lleaves for LightGBM. However, netsDB is slower than most of the GPU platforms in that case due to the increased GPU utilization. Compared to GPU platforms, the CPU platforms can reduce the monetary costs for models with $10$ trees and $500$ trees by $99\%$ and $44\%$, respectively. However, for serving models with $1,600$ trees, GPU can achieve better overall costs.


\subsubsection{Summary of Findings}
The key findings are:

\noindent
(1) On the CPU platforms, the in-database analytics outperformed other platforms in most cases, except that it is slightly worse than lleaves for large-scale model sizes (e.g., $1,600$ trees). That is because, when model size decreases, the data loading process becomes a more severe performance bottleneck.

\noindent
(2) GPU platforms outperformed netsDB on TPCx-AI using large models and Higgs and Airline using medium to large models. 
However, netsDB outperformed most GPU platforms in all other cases. 

\noindent
(3) NetsDB-Rel significantly improved the performance of medium to large-scale decision forest models compared to the netsDB-UDF. That is because model parallelism reduced the memory footprint and cache misses compared to data parallelism.

\noindent
(4) Lleaves, which compiles decision trees to partitioned and vectorized functions that consist of nested \texttt{if-else} blocks, achieved the best performance among CPU platforms for the LightGBM models.

\subsection{Wide and/or Sparse Datasets}
We also found that many popular datasets are wide and/or sparse. For example, in Bosch, Epsilon, and Criteo, each tuple has $968$, $2,000$, and $1$M features, respectively. Among these datasets, Bosch and Criteo are sparse datasets that contain missing values. 
The overall results are illustrated in Tab.~\ref{tab:bosch}, Tab.~\ref{tab:epsilon}, and Tab.~\ref{tab:crieto}, which are explained in detail as follows.

\begin{table*}[]
\begin{center}
\small
\resizebox{\textwidth}{!}{
\begin{tabular}{|c||c|c|c|c|c|c|c|c|c|c||c|c|c|c|}
\hline
 &\multicolumn{10}{|c||}{CPU} & \multicolumn{4}{c|}{GPU}\\
&Sklearn&ONNX&HB-Pytorch&HB-TS&HB-TVM&TreeLite&lleaves&netsDB-UDF&netsDB-Rel&netsDB-OPT&HB-Pytorch&HB-TS&HB-TVM&FIL\\ 
\hline
\multicolumn{15}{|c|}{\texttt{XGBoost}} \\ 
\hline
10 Trees  &22.8&23.0&22.6&23.7&23.8&23.6&-&\textbf{0.6}&4.2&2.4&\textbf{19.0}&20.0&19.5&19.6\\
500 Trees  &23.4&24.9&28.4&27.1&26.1&26.2&-&\textbf{3.4}&8.0&6.2&\textbf{19.4}&19.9&19.6&19.7\\
1600 Trees  &28.2&28.4&48.1&46.5&31.8&32.1&-&11.6&11.7&\textbf{10.1}&20.6&20.9&19.8&\textbf{19.7}\\
\hline
\multicolumn{15}{|c|}{\texttt{LightGBM}} \\ 
\hline
10 Trees  &22.9&22.6&22.9&23.2&23.2&23.3&23.3&\textbf{0.6}&4.2&2.4&\textbf{19.0}&19.9&19.5&19.9\\
500 Trees  &26.1&24.5&28.2&27.3&26.9&24.4&24.5&\textbf{3.2}&8.8&7.1&\textbf{19.4}&20.3&19.5&20.0\\
1600 Trees  &35.3&29.1&43.8&44.2&32.7&29.7&27.1&9.8&11.9&\textbf{9.5}&20.6&20.9&\textbf{19.8}&20.1\\
\hline
\end{tabular}
}
\caption{End-to-End Latency Comparison for the Bosch. (Unit: seconds)}
\label{tab:bosch}
\end{center}
\end{table*}


\begin{table*}[]
\begin{center}
\small
\resizebox{\textwidth}{!}{
\begin{tabular}{|c||c|c|c|c|c|c|c|c|c||c|c|c|c|}
\hline
 &\multicolumn{9}{|c||}{CPU} & \multicolumn{4}{c|}{GPU}\\
&Sklearn&ONNX&HB-Pytorch&HB-TS&HB-TVM&TreeLite&lleaves&netsDB-UDF&netsDB-Rel&HB-Pytorch&HB-TS&HB-TVM&FIL\\ 
\hline
\multicolumn{14}{|c|}{\texttt{RandomForest}} \\ 
\hline
10 Trees  &133.6&132.2&132.5&135.2&132.8&141.7&-&\textbf{0.4}&4.0&131.0&131.5&131.3&131.8\\
500 Trees  &135.5&135.4&137.7&136.5&135.0&142.4&-&\textbf{3.1}&5.7&131.2&131.7&131.5&131.8\\
1600 Trees  &138.1&135.1&141.8&141.6&135.9&-&-&9.7&\textbf{9.2}&131.6&132.1&131.5&131.9\\
\hline
\multicolumn{14}{|c|}{\texttt{XGBoost}} \\ 
\hline
10 Trees  &132.3&132.3&132.8&132.4&134.4&132.6&-&\textbf{0.5}&3.9&131.0&131.9&131.3&131.5\\
500 Trees  &132.6&135.2&135.4&141.1&135.7&134.6&-&\textbf{3.1}&5.7&131.2&132.4&131.5&131.6\\
1600 Trees  &133.1&135.0&140.5&139.4&137.2&136.6&-&9.6&\textbf{9.0}&131.6&132.6&131.6&131.6\\
\hline
\multicolumn{14}{|c|}{\texttt{LightGBM}} \\ 
\hline
10 Trees  &132.7&132.4&132.2&133.9&133.6&133.0&132.9&\textbf{0.5}&3.9&131.0&131.9&131.3&131.8\\
500 Trees  &134.0&134.0&135.0&135.7&134.3&134.2&135.0&\textbf{3.1}&5.4&131.2&132.1&131.5&131.8\\
1600 Trees  &136.2&135.2&139.2&137.4&135.8&143.1&-&9.1&\textbf{8.6}&131.6&132.3&133.4&132.0\\
\hline
\end{tabular}
}
\caption{End-to-End Latency Comparison for Epsilon. (Unit: seconds)}
\label{tab:epsilon}
\end{center}
\end{table*}

\begin{table}[]
\begin{center}
\small
\begin{tabular}{|c|c|c|c|}
\hline
&Sklearn&TreeLite&netsDB-UDF\\ 
\hline
\multicolumn{4}{|c|}{\texttt{RandomForest}} \\ 
\hline
10 Trees  &130.8&124.7&\textbf{2.2}\\
500 Trees  &409.0&152.1&\textbf{79.4}\\
1600 Trees  &1061.7&\textbf{216.3}&277.9\\
\hline
\multicolumn{4}{|c|}{\texttt{XGBoost}} \\ 
\hline
10 Trees  &125.2&126.2&\textbf{3.99}\\
500 Trees  &209.8&\textbf{191.9}&193.1\\
1600 Trees  &412.3&\textbf{326.7}&642.2\\
\hline
\multicolumn{4}{|c|}{\texttt{LightGBM}} \\ 
\hline
10 Trees  &132.0&126.6&\textbf{4.0}\\
500 Trees  &290.6&\textbf{141.7}&172.3\\
1600 Trees  &645.7&\textbf{216.2}&564.6\\
\hline
\end{tabular}
\caption{End-to-End Latency for Criteo.  (Unit: seconds)}
\label{tab:crieto}
\end{center}
\end{table}

\subsubsection{Bosch}
Bosch data in PostgreSQL contains NULL values. At the inference time, it is loaded from PostgreSQL as a Pandas DataFrame that contains nan values. Such sparse data is not directly supported in Scikit-learn for training a RandomForest model, which is needed by other dedicated inference platforms. Therefore, the RandomForest results are unavailable for this workload in Tab.~\ref{tab:bosch}. 

For models trained on top of sparse data, each tree node specifies a default branch to follow if the feature of the tree node is missing in the input sample. 

As illustrated in Tab.~\ref{tab:bosch}, for ten trees, netsDB-UDF achieved $37\times$ and $31\times$ speedup compared to the second-best CPU platform and the best GPU platform, respectively. For $500$ trees, netsDB-UDF achieved $8\times$ speedup on CPU and $6\times$ speedup on GPU. For $1,600$ trees, netsDB-OPT has better performance than netsDB-UDF, achieving $2.8\times$ speedup on CPU and $2.1\times$ speedup on GPU.

We found that wide and short datasets were inferred much faster than narrow and tall datasets of similar sizes. For example, the total size of Bosch is similar to Airline; however, the inference time of Bosch is significantly lower than Airline. That is because the number of tuples in Bosch is only $1\%$ of the number of tuples in the Airline dataset. (The computational complexity of the decision forest inference workload is mainly impacted by the number of tuples in the inference dataset, the number of trees, and the number of nodes in each tree.) 

Moreover, because Bosch and Airline datasets have similar sizes, their data transfer overheads are similar. As a result, for Bosch, the ratio of the data transfer latency to the end-to-end latency is significantly higher than the Airline workload. It explains the increased speedup of netsDB on the Bosch dataset compared to the Airline dataset. It also indicates that netsDB can achieve better speedup compared to other platforms for wide and short datasets.

\begin{figure}[h]
\centering
\vspace{-10pt}
\includegraphics[width=0.45\textwidth]{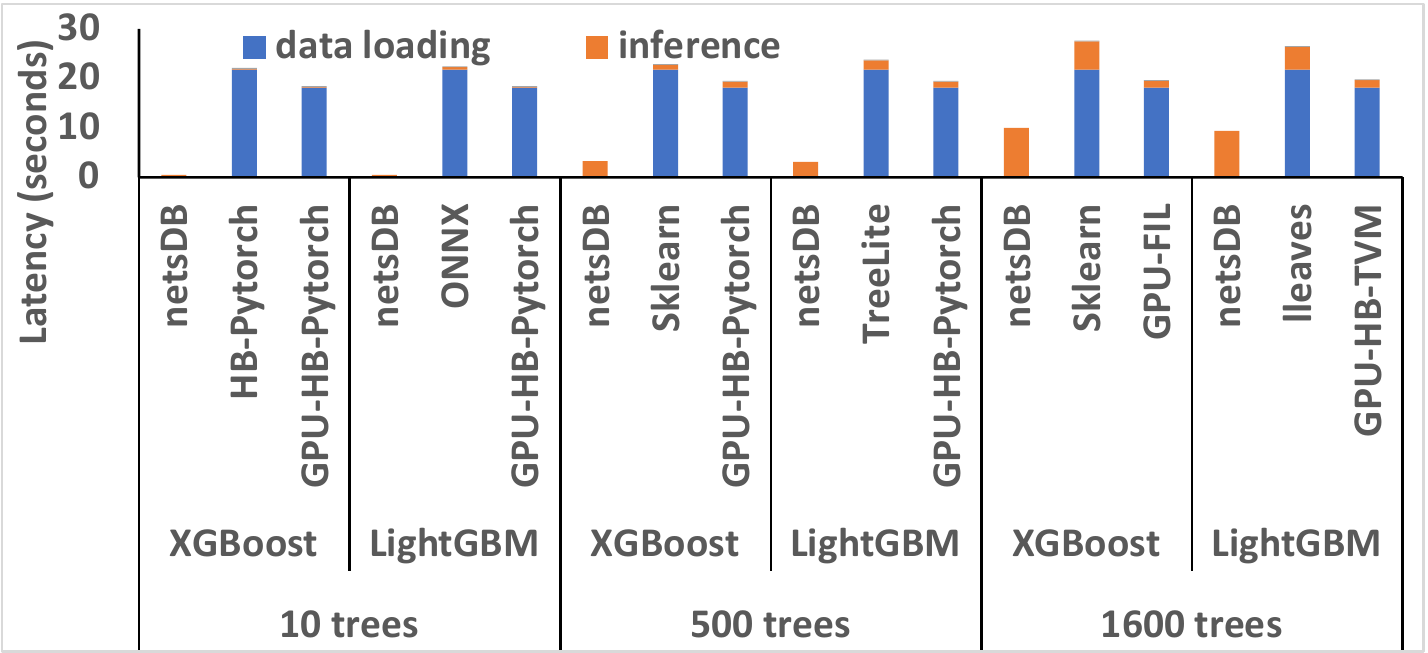}
\caption{\label{fig:bosch-breakdown} \small
{\color{black} Comparison of Latency breakdown on netsDB, the second-best CPU platform, and the best GPU platform for Bosch}
}
\end{figure}

\subsubsection{Epsilon}
\label{sec:epsilon}
As aforementioned, Epsilon is a wide and dense dataset with $2,000$ features. Because PostgreSQL does not support more than $1,600$ columns, we store each sample as one column on PostgreSQL in array type for all platforms except netsDB. In netsDB, the dataset is stored as a collection of tensor blocks in pages, and the page size can be flexibly configured.
%
As illustrated in Tab.~\ref{tab:epsilon}, netsDB achieved more than $300\times$, $40\times$, $10\times$ speedup for models with $10$, $500$, and $1600$ trees, respectively, compared to the fastest of the rest of CPU platforms and all GPU platforms. There are two reasons contributing to such huge performance gains. First, it turns out that it is expensive to convert a PostgreSQL array type back to a NumPy array, which becomes the bottleneck at the inference time in all platforms except netsDB. Second, this dataset contains fewer tuples than other datasets investigated in the paper. Therefore, the inference complexity is significantly lower than other narrower workloads of similar sizes (e.g., Higgs), given the same number of trees and the same depth of each tree.

\begin{figure*}[]
\centering
\vspace{-10pt}
\includegraphics[width=1\textwidth]{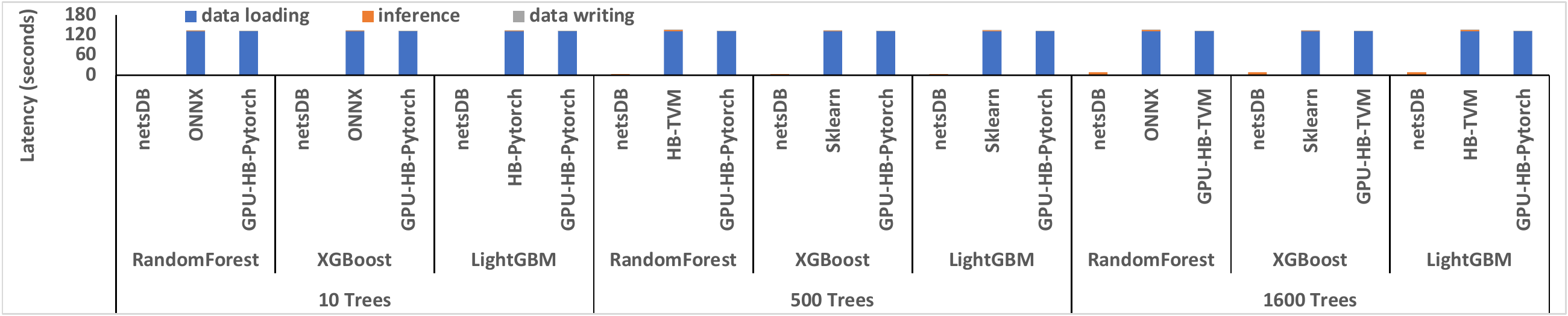}
\caption{\label{fig:epsilon-breakdown} \small
{\color{black} Comparison of Latency breakdown on netsDB, the second best CPU platform, and the best GPU platform for Epsilon}
}
\end{figure*}

As illustrated in Fig.~\ref{fig:epsilon-breakdown}, $99\%$ of time is spent in data loading (including converting the data received from the PostgreSQL array column into a NumPy array). This explained (1) the performance benefits brought by the in-database inference in netsDB compared to other platforms; (2)why all other platforms have similar latency (i.e., the data loading overhead, which becomes the bottleneck, is similar for all platforms except netsDB).

\subsubsection{Criteo}
\label{sec:criteo}
Criteo is stored in LIBSVM file format~\cite{libsvmdata, CC01a}, where each row contains a row index and a list of <column-index, non-zero-value> pairs, which reduced $80\%$ of storage space. Among the platforms studied in this work, sparse storage formats such as LIBSVM are only supported by Scikit-learn, TreeLite, and netsDB. Therefore, results for other platforms are unavailable. (Using dense format with other platforms failed due to out-of-memory errors.) 

According to our observation, using such a sparse format significantly reduced the end-to-end latency by reducing the data transfer overheads and the memory footprint. However, as a side effect, the ratio of the data transfer latency to the overall latency is also significantly reduced. As a result, less performance gain can be achieved via in-database inference compared to other workloads that used the dense storage format. As illustrated in Tab.~\ref{tab:crieto}, although netsDB-UDF achieved $30\times$ to $60\times$ speedup for $10$ trees compared to Sklearn and TreeLite, the performance of netsDB-UDF is significantly worse than TreeLite for $500$ and $1,600$ trees. 

\subsubsection{Summary of Findings}
The key takeaways are:

\noindent
(1) The in-database inferences represented by netsDB outperformed all CPU/GPU platforms if the wide dataset is stored in dense format with missing values represented as nan. That is because the data loading process is more of a performance bottleneck for dense datasets that are \textit{short-and-wide} than \textit{tall-and-narrow} datasets of similar sizes.

\noindent
(2) GPU is less helpful for the inferences over short and wide datasets than tall and narrow datasets of similar sizes. That is because the short and wide datasets contain fewer testing samples, reducing the workload's computational complexity.

\noindent
(3) In-database inferences significantly outperformed dedicated ML platforms if the data loading process involves a complicated transformation between the source format in data stores and the target format in ML platforms.

\noindent
(4) If a large dataset is extremely sparse (e.g., Criteo), a sparse storage format will significantly reduce the storage costs. However, the performance benefits of in-database inferences will diminish correspondingly. That is because the dataset becomes significantly smaller, which incurs less overhead for data transfer.

\subsection{Model Conversion}
\label{sec:model-conversion}

It is necessary to convert Scikit-learn models to other dedicated inference platforms such as ONNX, HummingBird, netsDB, TreeLite, and lleaves. Then the converted models must be loaded into corresponding platforms before running any inference tasks. 

In the results presented in previous sections, we did not consider the model conversion time and the model loading time because these are one-time costs and can be amortized to all inference tasks that use the same model. Here, we discuss the model conversion and loading overheads.

\begin{figure}[h]
\centering
\vspace{-10pt}
\subfigure[Model conversion (Unit: seconds)]{
\label{fig:conversion}
\includegraphics[width=0.48\textwidth]{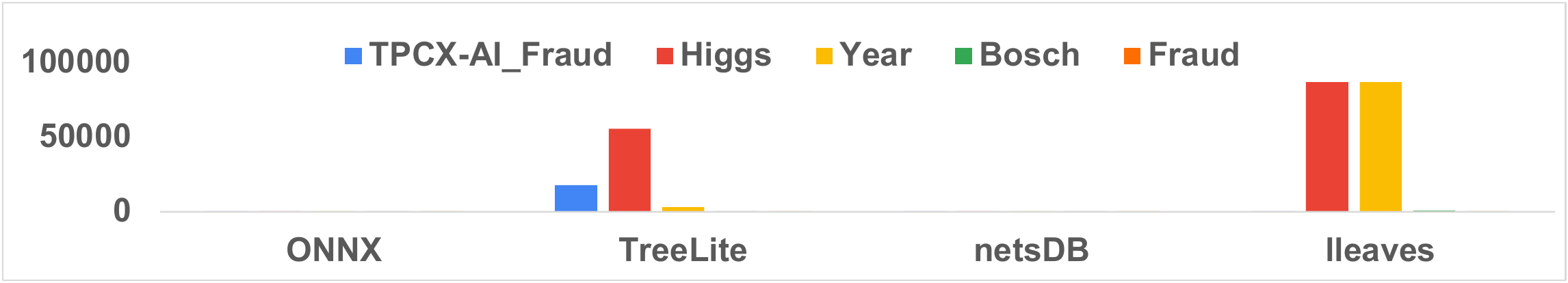}}
\subfigure[Model loading (Unit: milli-seconds)]{
\label{fig:loading}
\includegraphics[width=0.48\textwidth]{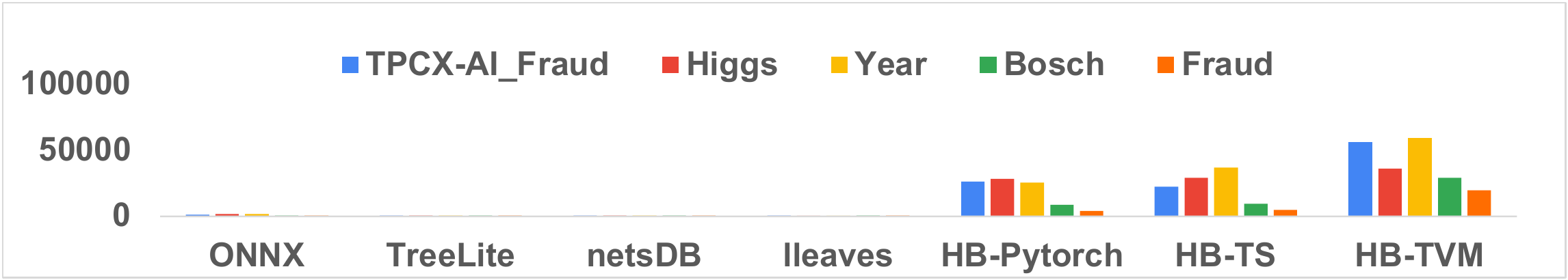}}
\caption{\label{fig:one-time-overheads} \small
{\color{black} Comparison of model conversion and loading overheads for 1600-tree LightGBM models. }
}
\end{figure}

As illustrated in Fig.~\ref{fig:conversion}, the platforms using the compiled tree traversal algorithm, such as TreeLite and lleaves, need more than one day to convert a $1,600$-tree model, which may hinder the adoption of such platforms. The conversion overheads for other platforms are around tens of seconds.

As illustrated in Fig.~\ref{fig:loading}, most of the converted models can be loaded in a short time that can be neglected. However, because the converted HummingBird model cannot be persisted, we convert the model during the loading process. That is why  HummingBird's loading process took significantly more time than other platforms.

\section{Detailed Analysis}
\label{sec:analysis}
In this section, we summarize and discuss the impacts of the various performance factors.

\noindent 
\textbf{Single-Thread Comparison}
%
TFDF is the only framework that implements the Quick-Scorer algorithm. However, it does not support multiple threads for inferences; and it incurs significant overheads when invoking the underlying C++-based decision forest library, called Yggdrasil~\cite{guillame2022yggdrasil}, by copying data multiple times. Therefore, we did not consider it in Sec.~\ref{sec:overall-evaluation} for fairness. Instead, we compared its performance of XGBoost to various CPU platforms here, all using a single thread. As illustrated in Tab.~\ref{tab:single-thread}, the results showed that QuickScorer achieved the best inference latency in the single-thread setting. Note that only inference time is measured, and the maximum tree depth is $6$ due to the leaf number limit in the QuickScorer implementation, as aforementioned in Sec.~\ref{sec:db-decisions}. 

\begin{table}[]
\begin{center}
\small
\begin{tabular}{|c||c|c|c|c|c|c|}
\hline
\#trees&Yggdrasil&TFDF&Sklearn&ONNX&TreeLite&HB-TVM\\ 
\hline
\multicolumn{7}{|c|}{\texttt{Higgs}} \\ 
\hline
10  &\textbf{0.7}&2.1&1.3&1.3&\textbf{0.7}&1.9\\
500  &\textbf{29.0}&45.4&31.9&34.8&87.6&98.4\\
1600  &\textbf{96.5}&140.5&98.9&105.0&340.5&298.1\\
\hline
\multicolumn{7}{|c|}{\texttt{Fraud}} \\ 
\hline
10    &\textbf{0.01}&0.31&0.03&0.02&\textbf{0.01}&0.03\\
500   &\textbf{0.26}&0.63&0.87&0.73&1.34&0.51\\
1600  &\textbf{0.48}&1.01&1.67&1.81&3.12&1.44\\
\hline
\multicolumn{7}{|c|}{\texttt{Year}} \\ 
\hline
10    &\textbf{0.002}&0.23&0.12&0.06&0.05&0.11\\
500  &1.27&2.86&1.58&1.60&4.08&\textbf{0.90}\\
1600  &\textbf{4.14}&7.08&4.93&4.87&17.80&2.61\\
\hline
\end{tabular}
\caption{\small XGBoost inference time using single thread.(Unit:seconds)}
\label{tab:single-thread}
\end{center}
\end{table}

\noindent 
\textbf{Further Discussions about CPU Platforms.} When using multi-threading, as illustrated in Sec.~\ref{sec:overall-evaluation}, 
Lleaves, which adopts the compilation-based algorithm and only supports LightGBM, achieved the best performance among most CPU platforms for LightGBM for large-scale models and large-scale inference data. Although both TreeLite and lleaves use similar code-generation ideas, lleaves significantly outperformed TreeLite because it employed LLVM and used more optimization techniques such as vectorization~\cite{lleaves}. 

\noindent 
\textbf{Further Discussions about GPU Platforms.} On GPU, the latest version of FIL achieved similar performance with HB-TVM (i.e., HummingBird using TVM as the backend). FIL is slightly better than HB-TVM for large decision forest models. We observed that the memory consumption of FIL is significantly lower than HB-TVM because the tensors used for replacing tree traversal required larger storage space. 
In addition, the performance results of HummingBird using different backends vary significantly, which indicates that the performance of the same algorithm is significantly impacted by the implementation.

\noindent 
\textbf{Parallelism Model} When the model size increases to $500$ to $1,600$ trees, we found that model parallelism significantly outperformed data parallelism. For example, for $500$ and $1,600$ trees, netsDB-UDF, which uses data parallelism, is significantly slower than netsDB-Rel, which uses model parallelism, as explained in Sec.~\ref{sec:overall-evaluation}. 
Model parallelism is preferred for large-scale models because after partitioning the model, each thread only requires access to a subset of trees, which greatly improves the cache locality. This is also proved by our profiling of cache misses.

Platforms that use data parallelism could also benefit from model partitioning. For example, In Yggdrasil, the underlying library of TF-DF, its optimized tree traversal algorithm, each thread will process $k$ trees at each iteration until all trees finish inference on the data partition. It achieved better performance than the naive tree traversal algorithm. Lleaves used similar ideas.

\noindent 
\textbf{Batching}
On most of the platforms, the performance improved with the increase in batch size until the processing of the batch exceeded available (memory) resources. For example, for TPCx-AI, netsDB-Rel achieved the best performance if we partitioned the TPCx-AI dataset into five batches. For HummingBird platforms using Pytorch and TorchScript as backends on the CPU, the optimal batch size is around $1,000$, while the TVM backend usually achieved the best performance with a batch size of around $100,000$.

\noindent 
\textbf{Vectorization} In netsDB-UDF and netsDB-Rel, we tuned vectorization granularity, which is the number of sample blocks that can be processed by each operator and the number of samples in each block. The results showed that the number of samples in a block will more significantly impact the performance than the number of sample blocks. It showed that in addition to vectorizing the relational processing operators (i.e., atomic operators), vectorizing underlying UDFs is even more important.

\section{Related Works}
\label{sec:relatedWorks}

\noindent
\textbf{Decision Forest Benchmark.} There are several benchmark frameworks for boosting-based tree algorithms. Microsoft Fast Retraining~\cite{fast-training, fast-training-blog} focused on the training time and accuracy of the two boosting-based algorithms, LightGBM and XGBoost, on CPU and GPU. GBM-perf~\cite{gbm-perf} compared the classification of the Airline dataset using H2O, XGBoost, LightGBM, and Catboost. Microsoft further developed the LightGBM benchmark suite~\cite{lightgbm-bench} to provide tooling for comparing implementations of boosting-based algorithms for both training and inferencing. Nvidia gbm-bench~\cite{gbm-bench} made Fast Retraining more scriptable and added support for the CatBoosting and RandomForest algorithm to the benchmark framework. They also ensure that hyperparameters are consistent for each training framework~\cite{dorogush2018every}.
While we leveraged these benchmarks a lot, we found that they did not consider data management and end-to-end latency. None of the existing studies explored the key design decisions of the end-to-end inference pipeline.

\noindent
\textbf{Other Related Works.} 
%
Raven~\cite{karanasos2019extending} proposed two techniques to co-optimize decision tree inference and SQL queries. The first technique is to prune the decision tree based on the selection queries. The second technique is to inline single and simple decision trees into UDFs leveraging the Froid framework~\cite{ramachandra2017froid}. However, Raven executes ensemble tree inferences (e.g., random forest) in ONNX runtime.
Clipper~\cite{crankshaw2017clipper} proposed various techniques to improve the performance of model serving (including decision forest models), such as inference result caching, dynamic batching, and model selection.
Browne et al.~\cite{browne2019forest} proposed memory packing techniques and a novel tree traversal method to overcome the random memory access overheads. 
A recent work, TreeBeard~\cite{prasad2022treebeard}, proposed an optimizing compiler based on the Multi-level Intermediate Representation. 
It achieved significant speedup compared to baselines such as TreeLite, HummingBird, and XGBoost. 
%
%
%
Tahoe~\cite{xie2021tahoe} and Sharp~\cite{sharp2008implementing} optimized decision forest inferences for GPU, and Owaida et al.~\cite{owaida2017scalable} proposed a decision forest inference framework for FPGA. 

\vspace{-5pt}
\section{Conclusion}
Decision forest, including RandomForest and boosting-based algorithms (e.g., XGBoost and LightGBM), is widely used in a variety of applications such as finance, healthcare, marketing, and search ranking,  because of its robustness, scalability to large-scale datasets, model interpretability, etc. However, most existing benchmarks and evaluations did not consider the management of the inference data and the data transfer to/from external data stores, which significantly impacts the overall performance, as shown in this study.

In this work, we have conducted a comprehensive comparison study for the end-to-end inference performance of decision forest models at different scales on eight platforms ($15$ platforms if consider variance in hardware, backends, etc.). We implemented our own in-database decision forest inference solution on netsDB, using two representations: UDF-Centric and Relation-Centric.

Our study showed that in-database inferences will save significant data loading/conversion time for runtime inferences. This is particularly important for two broad classes of workloads, where data transfer becomes a significant bottleneck. The first type of workloads serves large-scale datasets using small to medium-scale forest models with tens to hundreds of trees. The second type of workloads infers small-scale or wide-and-short datasets using all-scale models. These workloads argue for an in-database inference solution such as netsDB.
However, for workloads that exploit large-scale models, where inference rather than data transfer becomes the major performance bottleneck, in-database inference may not be an ideal solution due to its sub-optimal inference performance. Therefore, improving in-database inference performance for large-scale models by integrating with high-performance libraries and hardware acceleration could be a helpful direction. 

We believe these observations shed some light on the integration of database and AI/ML and may benefit the design of future systems.

\begin{acks} 
ASU students Jiaqing Chen and Yuze Liao have contributed to the work at the early stages of the benchmark development.
\end{acks}


\bibliographystyle{ACM-Reference-Format}
\bibliography{refs}  

\end{document}